\documentclass[aps,prstab,twocolumn,superscriptaddress]{revtex4-1}
\usepackage{amssymb}
\usepackage{amssymb}
\usepackage{amssymb}
\usepackage{graphicx}
\usepackage{graphics}
\usepackage{amsmath}
\usepackage{placeins}
\usepackage{hyperref}

\begin{document}

\title{Development and high-power testing of an X-band dielectric-loaded power extractor}

\author{\firstname{Jiahang} \surname{Shao}}
\email{jshao@anl.gov}
\affiliation{Argonne National Laboratory, Lemont, IL 60439, USA}
\author{\firstname{Chunguang} \surname{Jing}}
\affiliation{Argonne National Laboratory, Lemont, IL 60439, USA}
\affiliation{Euclid Techlabs, Bolingbrook, IL 60440, USA}
\author{\firstname{Eric} \surname{Wisniewski}}
\affiliation{Argonne National Laboratory, Lemont, IL 60439, USA}
\author{\firstname{Gwanghui} \surname{Ha}}
\affiliation{Argonne National Laboratory, Lemont, IL 60439, USA}
\author{\\\firstname{Manoel} \surname{Conde}}
\affiliation{Argonne National Laboratory, Lemont, IL 60439, USA}
\author{\firstname{Wanming} \surname{Liu}}
\affiliation{Argonne National Laboratory, Lemont, IL 60439, USA}
\author{\firstname{John} \surname{Power}}
\affiliation{Argonne National Laboratory, Lemont, IL 60439, USA}
\author{\firstname{Lianmin} \surname{Zheng}}
\email{Visiting Argonne from Tsinghua University in Beijing.}
\affiliation{Argonne National Laboratory, Lemont, IL 60439, USA}

\date{\today}

\begin{abstract}
Dielectric loaded structures are promising candidates for use in the structure wakefield acceleration (SWFA) technique, for both the collinear wakefield and the two-beam acceleration (CWA and TBA respectively) approaches, due to their low fabrication cost, low rf losses, and the potential to withstand high gradient. A short pulse ($\leq$20~ns) TBA program is under development at the Argonne Wakefield Accelerator (AWA) facility where dielectric loaded structures are being used for both the power extractor/transfer structure (PETS) and the accelerator. In this study, an X-band 11.7~GHz dielectric PETS was developed and tested at the AWA facility to demonstrate high power wakefield generation.  The PETS was driven by a train of eight electron bunches separated by 769.2~ps (9~times of the X-band rf period) in order to achieve coherent wakefield superposition. A total train charge of 360~nC was passed through the PETS structure to generate $\sim$200~MW, $\sim$3~ns flat-top rf pulses without rf breakdown. A future experiment is being planned to increase the generated rf power to approximately $\sim$1~GW by optimizing the structure design and improving the drive beam quality.
\end{abstract}


\maketitle

\section{Introduction}
The Advanced Accelerator Concepts (AAC) field conducts long-term research for a future large-scale collider with substantially higher energy and significantly lower cost than those can be achieved with current accelerator technology~\cite{ColbyRoadmap2016,CrosANAR2017}. Over the last several decades four schemes have been intensely investigated. Two are laser-driven schemes:  laser wakefield acceleration~\cite{EsareyRMP2009,LeemansPRL2014,SteinkeNature2016} and dielectric laser acceleration~\cite{PeraltaNature2013,EnglandRMP2014,WoottonOL2016}; and two are beam-driven schemes: plasma wakefield acceleration~\cite{HoganPRL2005,LitosNature2014,CordeNature2015} and structure wakefield acceleration~\cite{WeiPRL1988,JingPRL2006,GaoFPRST2008,JingPRL2011,AndonianPRL2012,SergeyAPL2012,WeiJPP2012,OSheaNatureComm2016,DanPRL2016,JingRAST2016,GaoPRL2018,LekomtsevPRAB2018,JingNIMA2018,JiahangIPAC2018,XueyingPRL2019}.

In SWFA, a high charge drive beam traveling through a structure excites wakefields which are used to accelerate a low charge main beam, in either the same structure (collinear wakefield acceleration) or a parallel structure (two-beam acceleration)~\cite{CrosANAR2017}. Based on current understanding, the TBA approach is favored over CWA for a linear collider for two reasons.  First, the beam transportation is less challenging since the beamline lattices for the main beam and the drive beam can be separately optimized~\cite{CrosANAR2017}; and second, the independent accelerating/decelerating structures provide more flexibility for over-all machine optimization~\cite{JingRAST2016}. Currently, both the mature design of the Compact Linear Collider (CLIC)~\cite{CLICCDR} and the AAC design of the Argonne Flexible Linear Collider (AFLC)~\cite{WeiJPP2012,JingRAST2016} are based on the TBA approach. The major difference between the two approaches is that AFLC is designed with a shorter rf pulse (20~ns vs. 230~ns in CLIC) with the expectation of achieving a higher accelerating gradient (270~MV/m vs. 100~MV/m in CLIC).  This is based on the published evidence of the rf breakdown rate decreasing with pulse length~\cite{GrudievPRST2009}. After optimizing structure and beam parameters, the efficiency of AFLC with such short rf pulses is comparable to that of CLIC~\cite{WeiJPP2012,JingRAST2016}.

In order to improve the gradient, efficiency, and cost of a TBA collider, various novel accelerating/decelerating structures have been studied: metallic two-halves structure~\cite{DanPRL2016,MassimoPRAB2016,ZhaPRAB2017}, dielectric-loaded structure (DLS)~\cite{WeiPRL1988,JingPRL2006,GaoFPRST2008,JingPRL2011,OSheaNatureComm2016}, metameterial structure~\cite{XueyingPRST2015,HummeltPRL2016,XueyingPRL2019}, photonic band gap structure~\cite{EvgenyaPRL2005,EvgenyaPRST2005,EvgenyaPRL2016}, and many others~\cite{ZouJAP2001,JingNIMA2008,AndonianPRL2012,SatohPRAB2016,SmirnovNIMA2016,JiahangIPAC2018,CahillPRAB2018,LekomtsevPRAB2018,HoangPRL2018}. Among them, the DLS is a very attractive candidate for a TBA collider due of its simple geometry, low fabrication cost, high group velocity with reasonable shunt impedance, and potential to withstand GV/m gradient.

DLSs were first used for particle acceleration in the late 1940s~\cite{BruckJAP1947,HarvieNature1948}.  Since then, steady progress has been made including an increased frequency range from GHz to THz~\cite{WeiPRL1988,JingPRL2006,GaoFPRST2008,JingPRL2011,OSheaNatureComm2016,DanRSI2018}, comparable shunt impedance as the metallic disk-loaded structure by using low loss microwave ceramic materials~\cite{GaoFPRST2008,JingRAST2016}, tunability with a second layer of nonlinear ferroelectric~\cite{JingPRL2011}, higher order modes (HOMs) damping with segmented conducting boundaries~\cite{ChojnackiJAP1991}, multipacting suppression by external magnetic field~\cite{JingAPL2013,JingAPL2016}, and GV/m level gradient in rf breakdown tests~\cite{ThompsonPRL2008} as well as CWA experiments~\cite{OSheaNatureComm2016}, etc. Despite this progress, high power rf pulse generation beyond 100~MW with drive bunch train excitation in the TBA approach was yet to be demonstrated for DLS. Previous results are limited to 30-50~MW due to the drive beam limitations as well as the immature design and fabrication methods of the DLSs~\cite{GaoFPRST2008,JingRAST2016,JiahangIPAC2017}.

In this paper, we present the development of an X-band 11.7~GHz dielectric-loaded power extractor, including design optimization, simulation, fabrication, cold-test, and high power test. This work is an important step towards the realization of the short-pulse TBA-based future linear colliders. It provides useful experimental information on rf breakdown in the nanosecond regime. Moreover, the development method can also be applied to other power extractors in general. The paper is organized as follows: Sec.~\ref{cha2} presents the structure design and simulation; Sec.~\ref{cha3} introduces the structure fabrication and the cold-test; Sec.~\ref{cha4} provides the high power test setup, the experimental results, and the data analysis; Sec.~\ref{cha5} presents the future study towards gigawatt rf power generation; and Sec.~\ref{cha6} summarizes the study.

\section{\label{cha2}Design and simulation}
A power extractor/transfer structure consists of a decelerating structure and an rf coupler. In the decelerating structure, the drive beam transfers its energy via wakefields into the TM$_{01}$ mode of the structure. Depending on the group velocity of the mode, the radiation moves either forward or backward with respect to the drive beam~\cite{XueyingPRL2019}. An rf coupler then converts the TM$_{01}$ mode into the rectangular TE$_{10}$ mode which can be further transferred with rectangular waveguides into an accelerating structure to accelerate the main beam.

\begin{figure}[h!tbp]
	\includegraphics[width=7.5cm]{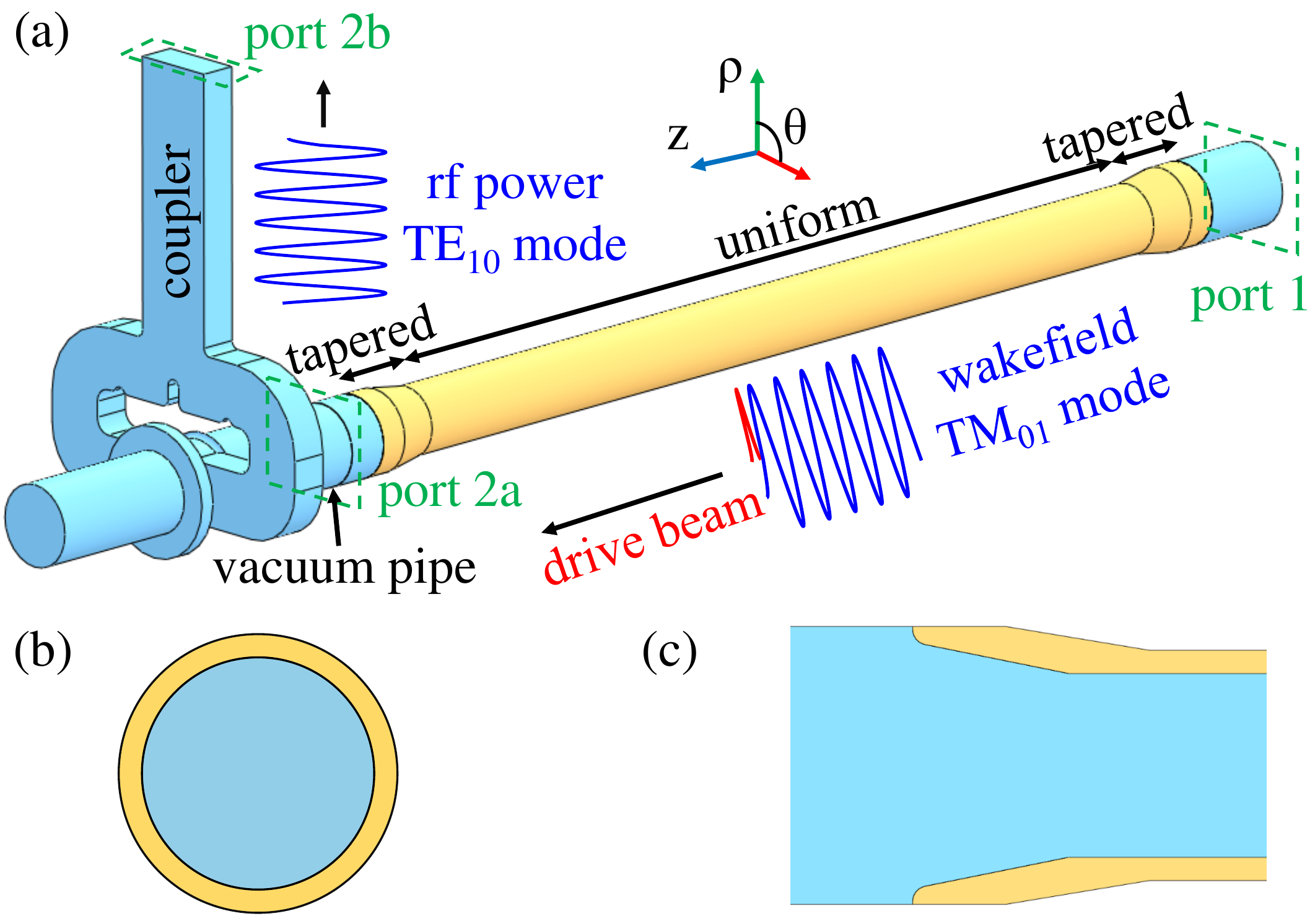}
	\caption{\label{Fig_dielectric_tube} A dielectric-loaded power extractor. The yellow and blue parts denote the dielectric and the vacuum respectively. The copper part which covers the outside of the whole structure is not shown. The drive beam travels through the center of the dielectric tube along the z-axis. (a) Layout of the whole structure. (b) Transverse cross section of the uniform dielectric section. (c) Longitudinal cross section of the tapered dielectric section.}
\end{figure}

The dielectric PETS used in this study (Fig.~\ref{Fig_dielectric_tube}) consists of a cylindrical DLS (uniform section) for deceleration followed by a matching section and an rf coupler to transfer the power into a rectangular waveguide. Deceleration of the drive beam takes place in the uniform section of the DLS (Fig.~\ref{Fig_dielectric_tube}b) where the beam-structure interaction is strongest.  Matching sections are located at both ends of the uniform section (Fig.~\ref{Fig_dielectric_tube}c) for impedance matching between the uniform section and the rf coupler. Note that, in general, the DLS can be designed in various forms, such as a uniform tube with constant impedance~\cite{GaoFPRST2008,JiahangIPAC2017}, a segmented tube with different impedance sections~\cite{JingIPAC2018}, etc.

In this section, the detailed design and simulation of the uniform section, the matching section, and the coupler will be presented respectively.

\subsection{Uniform section}
The design of the uniform section determines the strength and the duration of the generated rf power under certain drive beam conditions. In the design phase, the geometry is optimized based on various considerations including the required rf power level and pulse length, drive beam transportation, realistic dielectric material, and fabrication restrictions~\cite{GaoFPRST2008,JingRAST2016}. The design procedure can be divided into two parts: electromagnetic properties of the structure and rf pulse properties. The assumptions used in the derivation in this section are listed as follows. 1) Only the cylindrical TM$_{01}$ mode is taken into consideration for simplicity. 2) The electron beam is ultrarelativistic. 3) The longitudinal bunch profile has a Gaussian distribution.

\subsubsection{Electromagnetic Properties of the DLS}
Assuming a perfect outer metallic boundary (no rf losses), the propagation in the DLS is given by,
\begin{equation}\label{Eqn_propagation}
\left\{
\begin{aligned}
T_{1}^{2}+\beta ^{2}&=\omega_{0} ^{2} \epsilon _{0}\mu _{0}=\frac{\omega_{0} ^{2}}{c^{2}}\\
T_{2}^{2}+\beta ^{2}&=\omega_{0} ^{2} \epsilon _{0}\epsilon _{r}\mu _{0}\\
\end{aligned}
\right.
\end{equation}
where the subscript $1$ and $2$ denote the vacuum region and the dielectric region, respectively; $T_{1,2}$ is the cutoff wave number; $\beta$ is the propagation constant; $\epsilon _{0}$ is the vacuum permittivity; $\epsilon _{r}$ is the dielectric constant; $\mu _{0}$ is the vacuum permeability; $\omega_{0}$ is the frequency; and $c$ is the speed of light.

In the cylindrical TM$_{01}$ mode, $E_{\theta}=0$, $H_{\rho}=0$, and $H_{z}=0$. The other field components can be expressed as
\begin{equation}\label{Eqn_fieldcomponents}
\left\{
\begin{aligned}
E_{\rho,1}&=j\beta T_{1}AJ_{1}(T_{1}\rho)\\
E_{z,1}&=T_{1}^{2}AJ_{0}(T_{1}\rho)\\
H_{\theta,1}&=j\omega_{0} \epsilon _{0}T_{1}AJ_{1}(T_{1}\rho)\\
E_{\rho,2}&=j\beta T_{2}[BJ_{1}(T_{2}\rho)+CN_{1}(T_{2}\rho)]\\
E_{z,2}&=T_{2}^{2}[BJ_{0}(T_{2}\rho)+CN_{0}(T_{2}\rho)]\\
H_{\theta,2}&=j\omega_{0} \epsilon _{0}\epsilon _{r}T_{2}[BJ_{1}(T_{2}\rho)+CN_{1}(T_{2}\rho)]\\
\end{aligned}
\right.
\end{equation}
where $A$, $B$, and $C$ are constants; $J_{n}$ and $N_{n}$ are the $n$th-order Bessel functions of the first and the second kind respectively.

The boundary conditions are given by,
\begin{equation}\label{Eqn_boundarycondition}
\left\{
\begin{aligned}
E_{\rho,1}|_{\rho=a}&=E_{\rho,2}|_{\rho=a}\\
H_{\theta,1}|_{\rho=a}&=H_{\theta,2}|_{\rho=a}\\
E_{z,2}|_{\rho=b}&=0\\
\end{aligned}
\right.
\end{equation}
where $a$ and $b$ denote the inner and the outer radius of the tube respectively.

From the preceding equations, the dispersion relation can be derived as
\begin{equation}\label{Eqn_diespersion}
\frac{T_{2}J_{1}(T_{1}a)}{\epsilon _{r}T_{1}J_{0}(T_{1}a)}=\frac{J_{1}(T_{2}a)N_{0}(T_{2}b)-J_{0}(T_{2}b)N_{1}(T_{2}a)}{J_{0}(T_{2}a)N_{0}(T_{2}b)-J_{0}(T_{2}b)N_{0}(T_{2}a)}
\end{equation}

The phase velocity of the cylindrical TM$_{01}$ mode $v_{p}=\omega_{0} /\beta$ should be equal to the beam velocity for synchronization. This is accomplished by solving Eqn.~\ref{Eqn_propagation} and Eqn.~\ref{Eqn_diespersion} for $b$ given a set of $a$, $\epsilon _{r}$, and $\omega_{0}$.

Once $b$ is solved, the variation of $\beta$ as a function of $\omega_{0}$ can be derived from Eqn.~\ref{Eqn_propagation} and Eqn.~\ref{Eqn_diespersion} so as to calculate the group velocity $v_{g}\equiv \beta_{g}c=d\omega_{0} /d\beta$.

After considering a realistic copper boundary with rf losses, the stored energy $U$, the power loss on the outer copper wall $P_{wall}$, and the power loss in the dielectric $P_{diel}$ can be calculated as
\begin{equation}\label{Eqn_energy_loss}
\left\{
\begin{aligned}
U=&\frac{1}{2}\epsilon _{0}\int _{V_{1}}|E_{\rho,1}+E_{z,1}|^{2}dv\\
&+\frac{1}{2}\epsilon _{0}\epsilon _{r}\int _{V_{2}}|E_{\rho,2}+E_{z,2}|^{2}dv\\
P_{wall}=&\frac{1}{2\sigma \delta}\oint _{S}|H_{\theta,2}|_{\rho=b}|^{2}ds\\
P_{diel}=&\frac{\omega_{0} \epsilon _{0}\epsilon _{r}\tan(\delta _{d})}{2}\int _{V_{2}}|E_{\rho,2}+E_{z,2}|^{2}dv\\
\end{aligned}
\right.
\end{equation}
where $S$ denotes the surface between copper and dielectric; $V_{1}$ and $V_{2}$ denote the volume of the vacuum and the dielectric respectively; $\sigma$ is the conductivity of copper; $\delta = \sqrt{2/(\omega \mu \sigma)}$ is the skin depth; and $\tan (\delta _{d})$ is the loss tangent of the dielectric material.

Therefore, the quality factor $Q$ and the shunt impedance per unit length $r$ of the uniform section are given by,
\begin{equation}\label{Eqn_Q_roverQ}
\left\{
\begin{aligned}
Q=&\frac{\omega_{0} U}{P_{wall}+P_{diel}}\\
r=&\frac{(\int _{0}^{L_{s}}E_{z,1}|_{\rho=0}dl)^{2}}{(P_{wall}+P_{diel})L_{st}}\\
\end{aligned}
\right.
\end{equation}
where $L_{st}$ is the length of the uniform section.

\begin{figure}[h!tbp]
	\includegraphics[width=8.5cm]{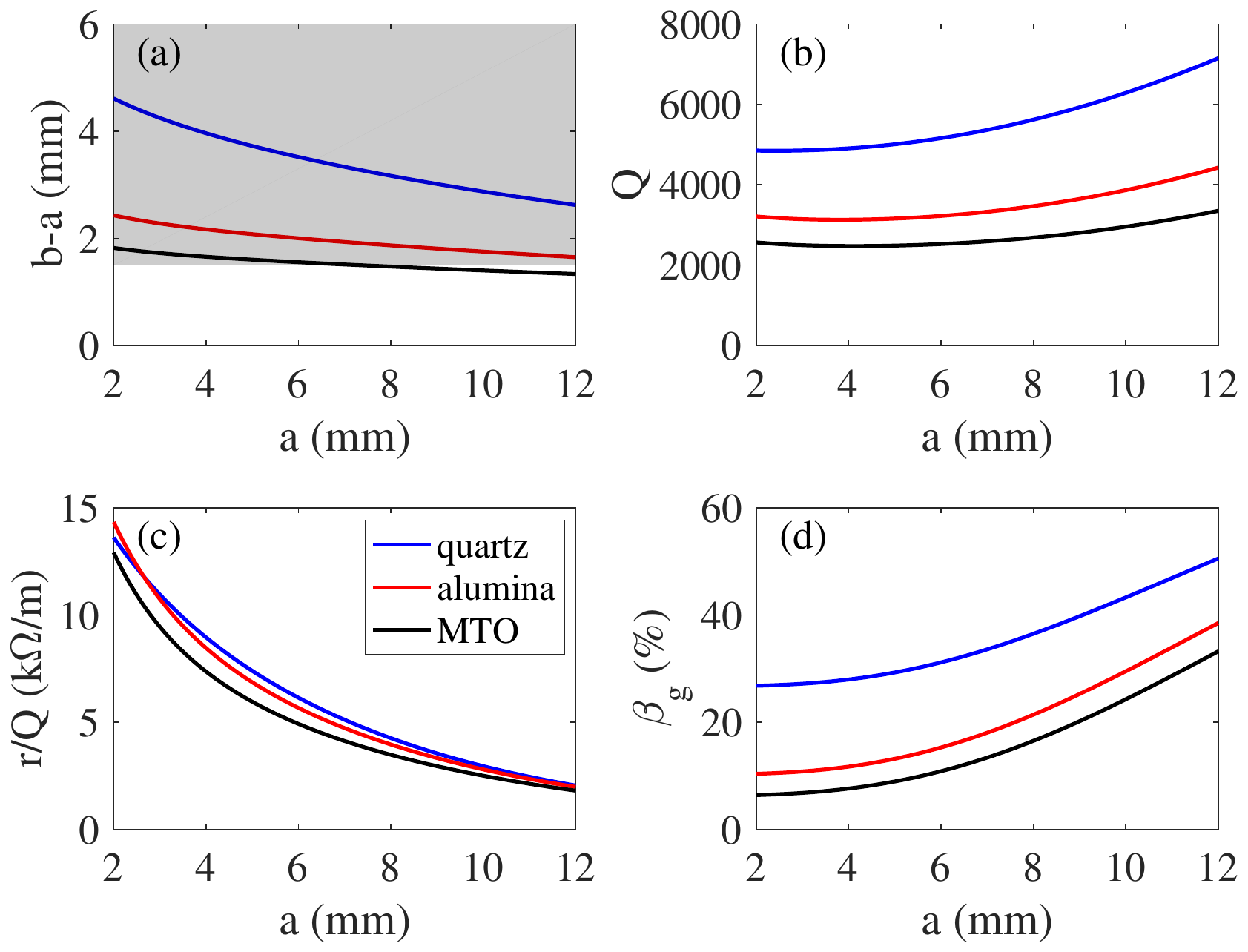}
	\caption{\label{Fig_uniform_electromagnetic} Electromagnetic properties of the uniform section as a function of the inner radius for different dielectric materials. (a) Dielectric wall thickness. (b) Quality factor. (c) $r/Q$. (d) Group velocity.}
\end{figure}

Three low loss dielectric materials with $\tan (\delta _{d})=1\times 10^{-4}$ were considered for the design: quartz ($\epsilon _{r}$=3.75), alumina ($\epsilon _{r}$=9.8), and MTO (MgTiO$_{3}$-Mg$_{2}$TiO$_{4}$, $\epsilon _{r}$=16)~\cite{kanareykinAAC2004}. When the power extractor operates at 11.7~GHz and is synchronized to an ultrarelativistic drive beam, the electromagnetic properties as a function of the inner radius have been numerically calculated, as illustrated in Fig~\ref{Fig_uniform_electromagnetic}.

\subsubsection{RF Power generated by the beam-driven DLS}
In this subsection, we first derive the rf parameters due to a single (subscript `s') drive bunch and then the rf parameters due to multiple (subscript `m') drive bunches (i.e. a drive bunch train) .   

When an ultrarelativistic drive bunch passes through the uniform section of the DLS ($v_{g}>0$), the leading edge of its wakefield packet travels together with the bunch while the trailing edge moves forward at the group velocity~\cite{GaoFPRST2008,JingRAST2016}. Therefore, the rf pulse length of the generated wakefield packet at the end of the DLS is given by
\begin{equation}\label{Eqn_pulselength_single}
\tau =\frac{L_{st}}{v_{g}}-\frac{L_{st}}{c}=\frac{L_{st}(1-\beta _{g})}{c\beta _{g}}
\end{equation}

The longitudinal electric field of the generated wakefield at the end of the DLS, including structure attenuation, is ($0\leq t \leq \tau$)
\begin{equation}\label{Eqn_field_single}
\begin{aligned}
E_{s}(t)=&|E_{s}(t)|\cos(\omega_{0} t)\\
=&2\Phi Q_{b}H(ct)\kappa _{l}e^{-\frac{\alpha c \beta _{g}t}{1-\beta _{g}}}\cos(\omega_{0} t)\\
\end{aligned}
\end{equation}
where $Q_{b}$ is the charge; $\Phi = \exp[-(\omega_{0} \sigma _{z}/c)^2/2]$ is the form factor; $\sigma _{z}$ is the root-mean-square (rms) bunch length; $\kappa _{l} = (\omega_{0} /4)(r/Q)[1/(1-\beta _{g})]$ is the loss factor per unit length~\cite{Danthesis2016}; $\alpha=\omega_{0} /(2Qv_{g})$ is the attenuation factor of the structure; and $H$ is the Heaviside step function. The time, t=0, is set when the drive beam reaches the exit of the uniform section.

The spectrum of the generated wakefield, given by the Fourier transform of Eqn.~\ref{Eqn_field_single} after ignoring structure attenuation, is
\begin{equation}\label{Eqn_spectrum_single}
F_{s}(\omega)=D \rm{sinc}[\frac{\tau (\omega -\omega_{0})}{2}]
\end{equation}
where $D$ is a constant.

The average energy loss of the bunch, by integrating the experienced field over the entire structure, is
\begin{equation}\label{Eqn_energyloss_single}
U_{loss}=\Phi \int _{0}^{L_{st}/c}cE_{s}(0)dt=\Phi ^{2}Q_{b}\kappa _{l}L_{st}
\end{equation}

The generated rf power at end the end of the DLS, following the general relation of $P=E^{2}v_{g}/(\omega r/Q)$ in traveling-wave structures, can be expressed as
\begin{equation}\label{Eqn_power_single}
P_{s}(t)=\frac{c\beta _{g}[E_{s}(t)]^{2}}{4\kappa _{l}(1-\beta _{g})}
\end{equation}

When a multi-bunch train of $N$ bunches separated by $T_{b}$ passes through the uniform section, the rf fields from each bunch superpose to give
\begin{equation}\label{Eqn_field_multi}
\begin{aligned}
E_{m}(t)=&\sum _{k=1}^{N} E_{s,k}[t-(k-1)T_{b}]\\
\equiv&|E_{m}(t)|\cos(\omega t)\\
\end{aligned}
\end{equation}
where $E_{s,k}$ is the longitudinal electric field excited by the $k$th bunch.

In this case, the spectrum of the generated wakefield, given by the Fourier transform of Eqn.~\ref{Eqn_field_multi} after ignoring structure attenuation, is
\begin{equation}\label{Eqn_spectrum_multi}
F_{m}(\omega)=\sum _{k=1}^{N} F_{s,k}(\omega)e^{-j(k-1)\omega T_{b}}
\end{equation}
where $F_{s,k}$ is the spectrum excited by the $k$th bunch.

The average energy loss of the $k$th bunch in the train, by integrating its experienced field over the entire structure, is
\begin{equation}\label{Eqn_energyloss_multi}
U_{loss,k}=\Phi \int _{0}^{L_{st}/c}cE_{m}[(k-1)T_{b}]dt
\end{equation}

Due the finite length of the rf pulse excited by a single bunch, $E_{m}$ reaches its maximum value after a rise time due to $N_{rise}=\lceil \tau /T_{b} \rceil -1$ bunches. For the rf pulse excited by $N$ bunches, the rise time $t_{rise}$, the flat-top duration $t_{flat}$, and the fall time $t_{fall}$ can be expressed as
\begin{equation}\label{Eqn_pulse_time}
\left\{
\begin{aligned}
t_{rise}=&N_{rise}T_{b}\\
t_{flat}=&(N-N_{rise})T_{b}\\
t_{fall}=&\tau -T_{b}\\
\end{aligned}
\right.
\end{equation}

When the drive bunches are spaced at integer wavelengths, $T_{b}=2\pi n/\omega_{0}$, the  $E_{s,k}$ add coherently to yield the maximum possible amplitude of the multi-bunch field
\begin{equation}\label{Eqn_field_multi_coherent}
|E_{m}(t)|=\sum _{k=1}^{N} |E_{s,k}[t-(k-1)T_{b}]|
\end{equation}

The average field amplitude of the flat-top can be expressed as
\begin{equation}\label{Eqn_field_flat_top}
\overline{E_{flat}}=\frac{\int _{N_{rise}T_{b}}^{NT_{b}}|E_{m}(t)|dt}{t_{flat}}\propto Q_{b}
\end{equation}

Therefore, the average power of the flat-top, following $P=E^{2}v_{g}/(\omega r/Q)$, is
\begin{equation}\label{Eqn_power_flat_top}
\overline{P_{flat}}=\frac{c\beta _{g}\overline{E_{flat}}^{2}}{4\kappa _{l}(1-\beta _{g})}\propto Q_{b}^{2}
\end{equation}

\begin{figure}[h!tbp]
	\includegraphics[width=8cm]{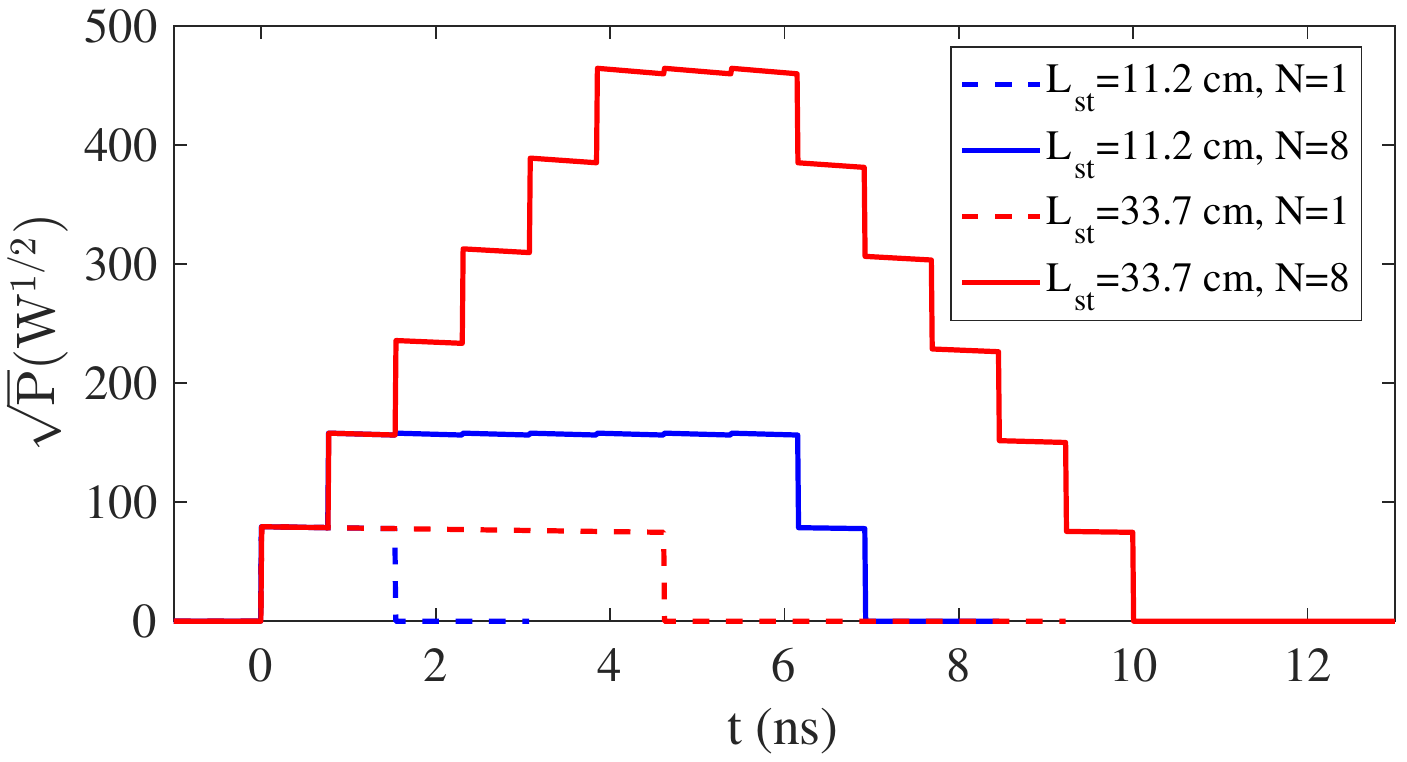}
	\caption{\label{Fig_port_signal_length} Calculated square root of the output power using the structure parameters listed in Table~\ref{Tab_uniform} but with various lengths. In the calculation, $T_{b}$=769.2~ps, $Q_{b}$=1~nC, and $\sigma _{z}$=1.5~mm.}
\end{figure}

For a given set of $T_{b}$, $N$, $Q_{b}$, and $\sigma _{z}$, $\overline{P_{flat}}$ can be increased by using longer structures at the expense of a shorter rf flat-top $t_{flat}$, as illustrated in Fig~\ref{Fig_port_signal_length}. Therefore, a long uniform section is chosen in this study in order to maximize the rf power level. Its length is fixed at 26~cm which is the practical limit of the current fabrication technology for a single dielectric tube. This length of dielectric tube restricts the wall thickness to be greater than  1.5~mm in order to prevent it from cracking during fabrication. This restriction is indicated as the shaded region in Fig.~\ref{Fig_uniform_electromagnetic}(a).

\begin{figure}[h!tbp]
	\includegraphics[width=8.5cm]{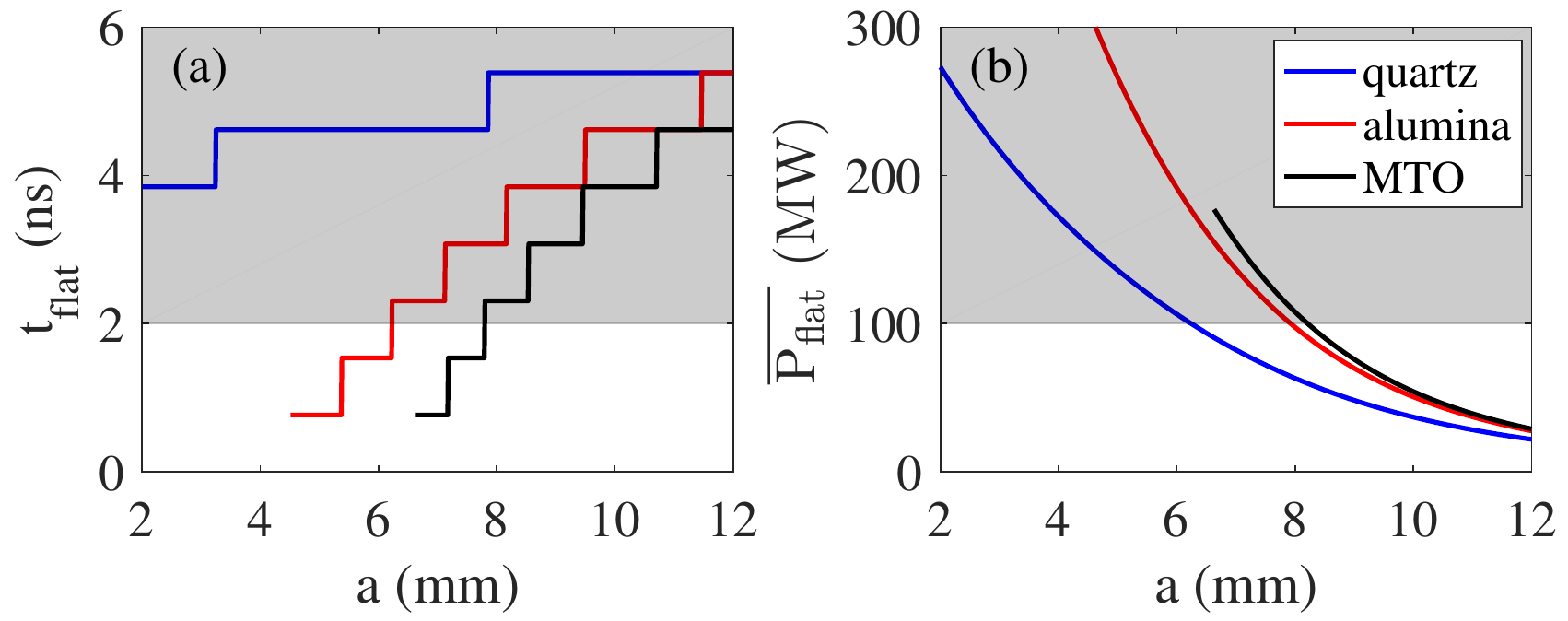}
	\caption{\label{Fig_uniform_wake} The duration (a) and the average power (b) of the flat-top as a function of the inner radius for different dielectric materials.}
\end{figure}

Given the fixed uniform section length of 26~cm, $t_{flat}$ and $\overline{P_{flat}}$ of the structures illustrated in Fig.~\ref{Fig_uniform_electromagnetic} can be numerically calculated based on the actual drive bunch parameters available at the Argonne Wakefield Accelerator facility: $T_{b}$=769.2~ps (1.3~GHz or the 9th subharmonic of the X-band frequency), $N$=8, and $Q_{b}\leq$60~nC~\cite{ManoelIPAC2017}. Figure~\ref{Fig_uniform_wake} illustrates the results for $Q_{b}$=30~nC and $\sigma _{z}$=~1.5~mm which are typical for routine operation. The design goals are to have a flat-top duration greater than 2~ns and an average power greater than 100~MW. The shaded regions indicate where the design goals are satisfied.

Accounting for both the electromagnetic and rf generation properties (Fig.~\ref{Fig_uniform_electromagnetic} and Fig.~\ref{Fig_uniform_wake}, respectively) of the DLS, we can now find the values of inner radius and materials that satisfy our design goals.  These are found to be: $a\leq$6.3~mm for quartz; 6.2~mm$\leq a\leq$7.9~mm for alumina; and no proper $a$ for MTO. Given the difficulty of transmitting a high charge drive bunch train through the DLS, a larger inner radius is preferred. Therefore, alumina has been chosen with $a$ set to 7.50~mm. The detailed parameters of the uniform section used in this study are listed in Table~\ref{Tab_uniform}.

\begin{table}[h!tbp]
	\caption{\label{Tab_uniform}Parameters of the uniform section.}
	\begin{ruledtabular}
		\begin{tabular}{l c}
			Parameter&Value\\
			\hline
			Material&Alumina\\
			Dielectric constant&9.8\\
			Loss tangent&$1 \times 10^{-4}$\\
			Inner radius&7.50~mm\\
			Outer radius&9.40~mm\\
			Length&26~cm\\
			Quality factor&3393\\
			$r/Q$&4.32~k$\Omega$/m\\
			Group velocity&0.20~c\\
			$E_{surface,max}/E_{axis}$&1\\
			Flat-top duration (8-bunch train)&3.1~ns\\			
		\end{tabular}
	\end{ruledtabular}
\end{table}

\subsection{Matching section and coupler}
An rf coupler is installed at the end of the uniform section to convert the TM$_{01}$ mode of the circular DLS into the TE$_{10}$ mode of the rectangular WR-90 waveguide. In previous studies of dielectric PETS, the rf output coupler was customized and could not be re-used for other structures~\cite{GaoFPRST2008,JiahangIPAC2017}. To simplify the design and to reduce the cost, a broad-band dual-feed coupler has been modified from the 11.424~GHz version broadly used in X-band metallic structure high gradient researches~\cite{NantistaPRST2004,DolgashevPAC2005}. 

\begin{figure}[h!tbp]
	\includegraphics[width=8cm]{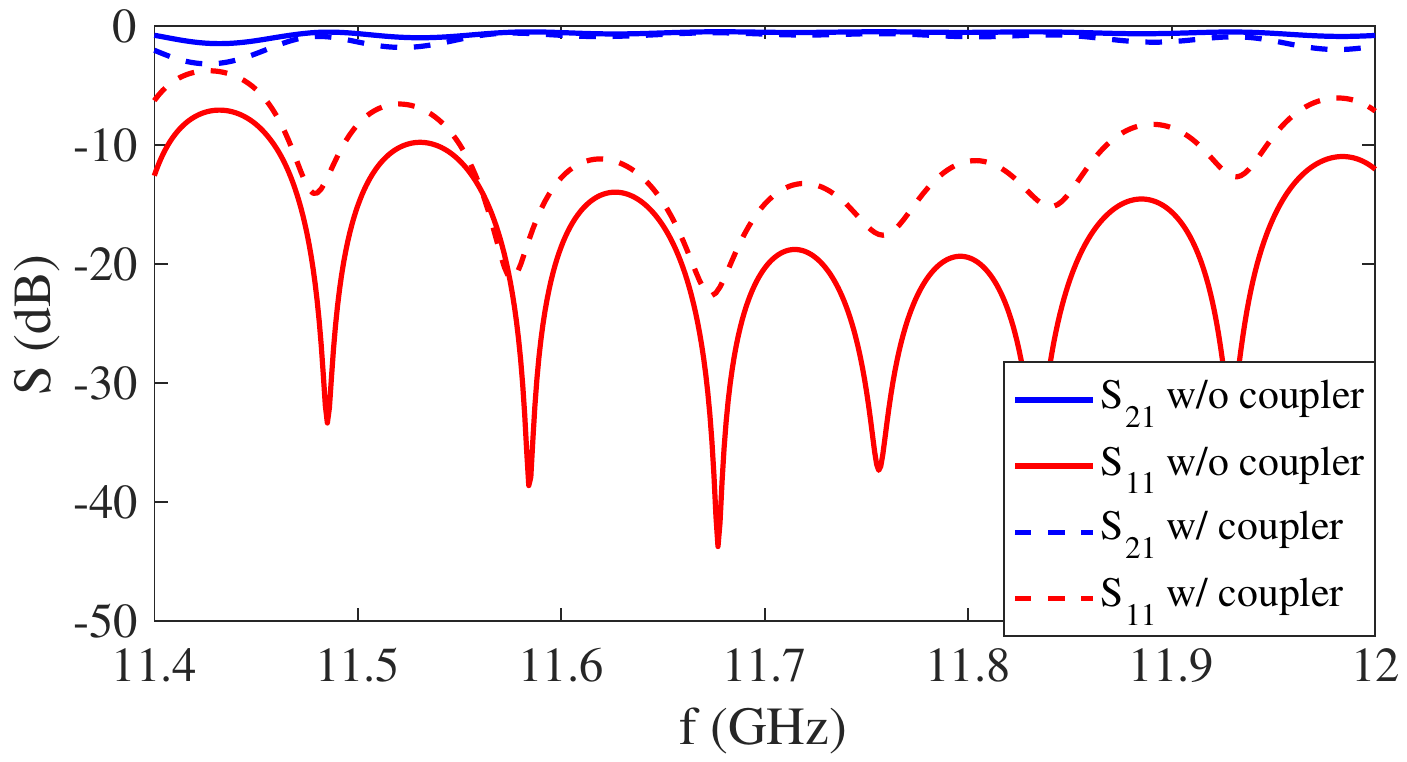}
	\caption{\label{Fig_simulation_S} Simulated S-parameters of the structure. The port definition is illustrated in Fig.~\ref{Fig_dielectric_tube}: Port 1 at the beam entrance; Port 2 at the beam exit when simulated without the coupler (Port 2a) and at the waveguide port when simulated with the coupler (Port 2b).}
\end{figure}

A dielectric matching section is used in order to match the impedance between the uniform dielectric section (inner diameter of 15.0~mm) and the rf coupler (vacuum pipe diameter of 22.7~mm) as illustrated in Fig.~\ref{Fig_dielectric_tube}(c).  An adiabatic matching section, with a gradual dielectric taper, is used to achieve a wide band width for short rf pulses. The optimal results were achieved by adjusting the length and angle of the dielectric tapered section, as illustrated in Fig.~\ref{Fig_simulation_S} (simulation with CST Microwave Studio~\cite{CSTmanual}). The results have also been verified with the ACE3P code~\cite{KwokLINAC2010}. For the entire structure, including the coupler, we achieved S$_{21}$ at 11.7~GHz of -0.7~dB with 3~dB bandwidth of over 1~GHz. The peaks separated by $\sim$90~MHz in the S$_{11}$ curve are caused by the matching section design. The total length of the DLS plus the matching sections at both ends is 30~cm. The length of the entire structure including the rf coupler is 42~cm.

\begin{figure}[h!tbp]
	\includegraphics[width=8cm]{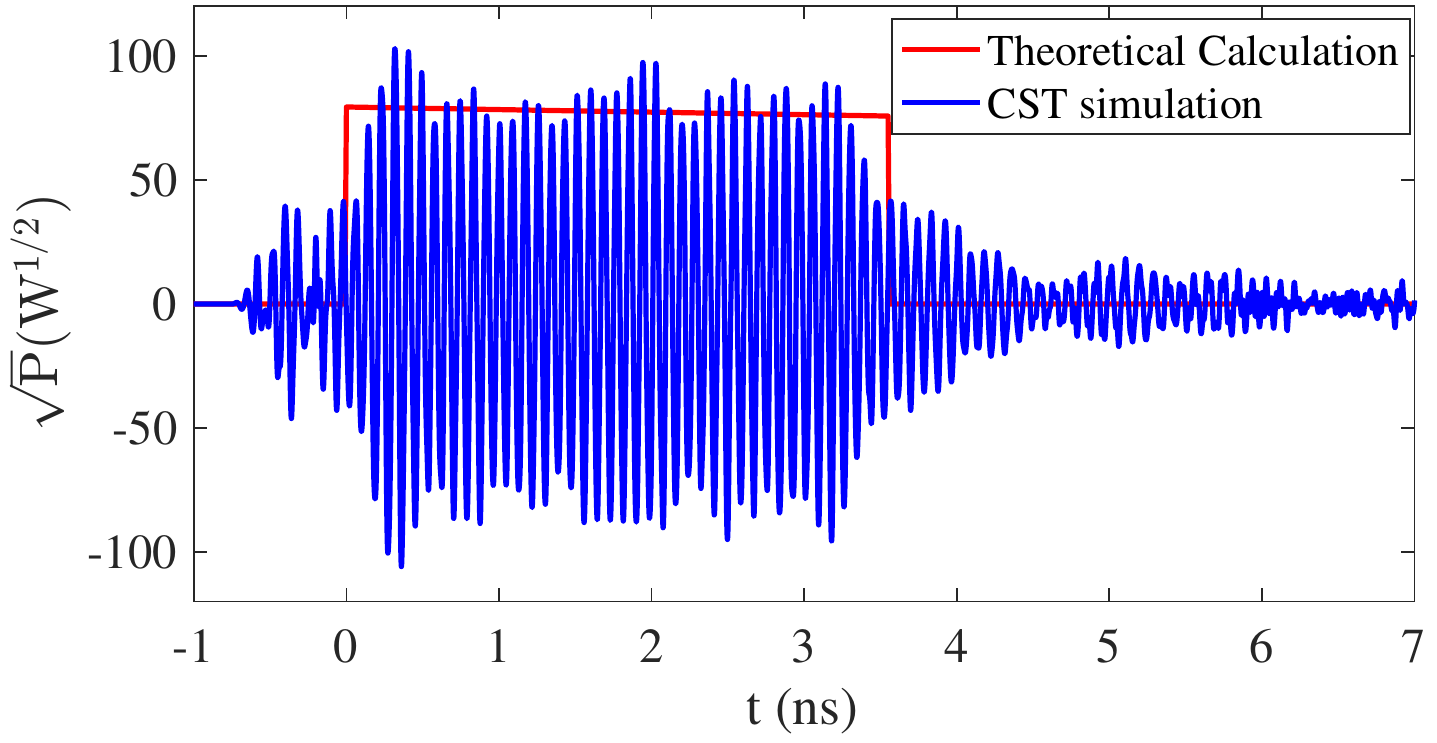}
	\caption{\label{Fig_port_signal} Coupler output signal driven by a single bunch with $Q_{b}$=1~nC and $\sigma _{z}$=1.5~mm. The time, t=0, is set at the start of the analytically calculated rf pulse from Eqn.~\ref{Eqn_power_single}}
\end{figure}

The rf output from the dielectric PETS, shown at the coupler port (Fig.~\ref{Fig_dielectric_tube}, port 2b) when driven by a single bunch, is shown in Fig.~\ref{Fig_port_signal} (simulation with CST Particle Studio, wakefield module~\cite{CSTmanual}). The length and amplitude are in good agreement with the analytic estimates based on Eqn.~\ref{Eqn_pulselength_single} and Eqn.~\ref{Eqn_power_single} which only take the uniform section into consideration. In the CST simulation, the rising time is caused by the finite bandwidth of the structure and the falling time is a result of the bandwidth as well as the transient process of the wakefield entering the downstream vacuum pipe~\cite{GrigorevaPRAB2018}. Finally, the non-ideal flatness is due to a small reflection arising from a slight impedance mismatch between the uniform section and the rf coupler.

\begin{figure}[h!tbp]
	\includegraphics[width=8cm]{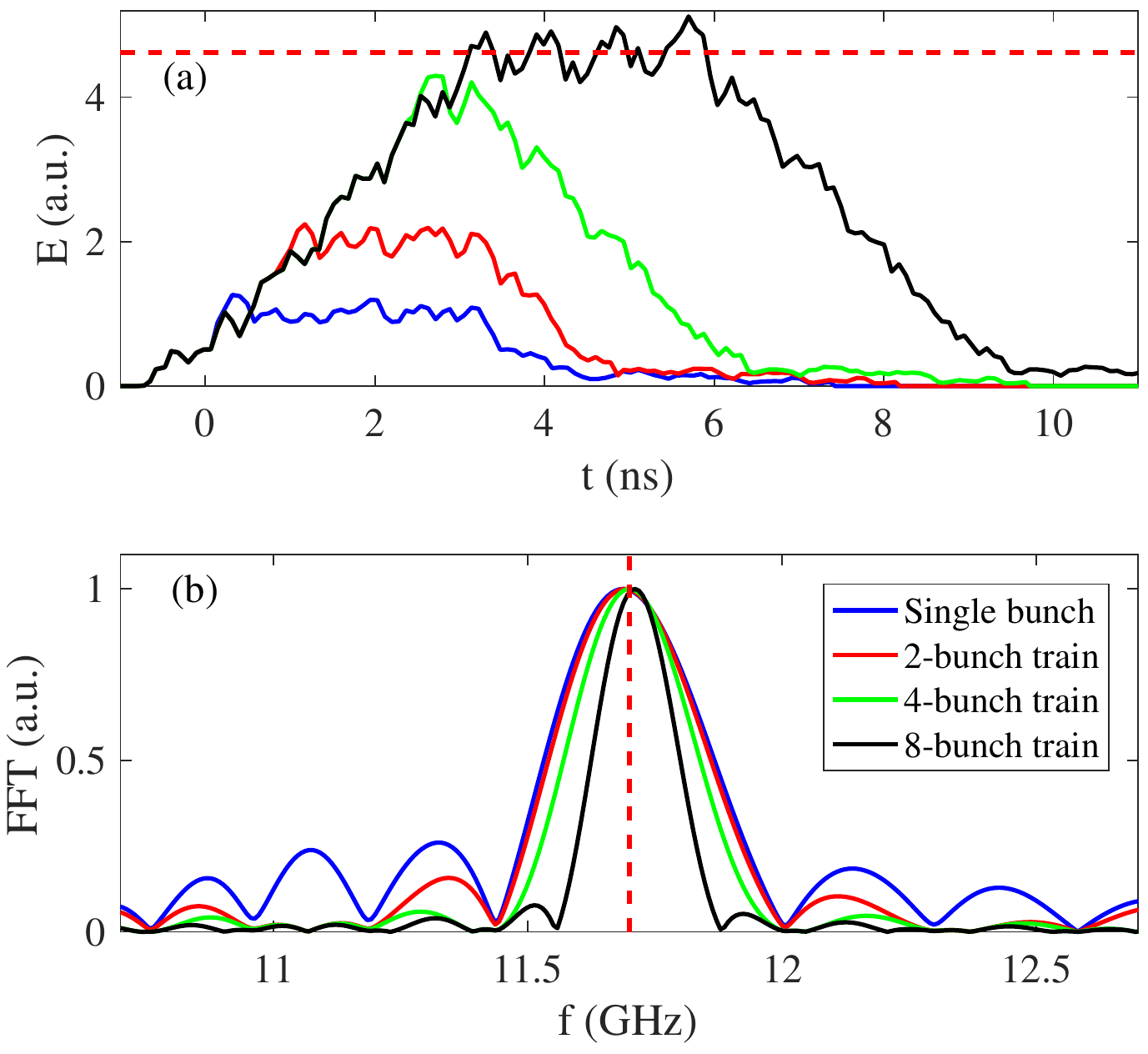}
	\caption{\label{Fig_port_signal_multi} The envelop (a) and the spectrum (b) of the coupler output rf signal. For bunch train cases, the field is calculated by adding up the ones driven by a single bunch with proper delays. The field amplitude is normalized to that driven by a single bunch. The horizontal and vertical red dashed lines indicate the average amplitude of the flat-top driven by the 8-bunch train and $f$=11.7~GHz, respectively.}
\end{figure} 

When driven by a multi-bunch train separated by 769.2~ps ($T_{b}=2\pi/\omega_{0} \times 9$), the port signal is coherently enhanced, as illustrated in Fig.~\ref{Fig_port_signal_multi}. The flat-top of the rf pulse is reached at the 5th drive bunch and has a duration of $\sim$3~ns. The average field amplitude and rf power of the flat-top driven by the 8-bunch train are 4.6 and 21.4~times of those driven by a single bunch. The full widths at half maximum (FWHM) of the frequency spectrum driven by a single bunch, a 2-bunch train, a 4-bunch train, and an 8-bunch train are 0.34~GHz, 0.32~GHz, 0.27~GHz, and 0.18~GHz respectively; all of which are in good agreement with those calculated by Eqn.~\ref{Eqn_spectrum_multi}.

\section{\label{cha3}Fabrication and cold-test}

\subsection{Fabrication}
\begin{figure}[h!tbp]
	\includegraphics[width=8cm]{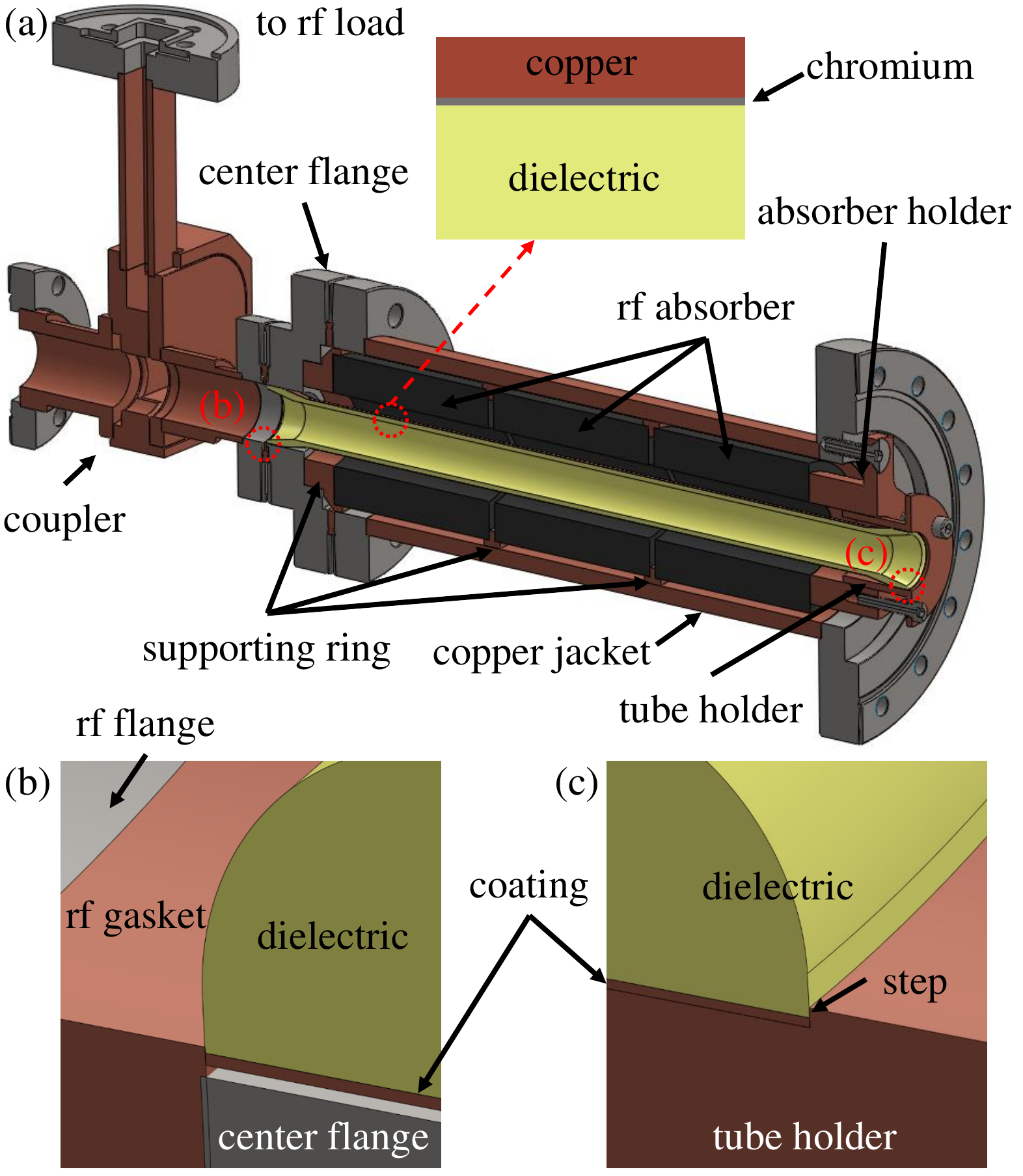}
	\caption{\label{Fig_whole_assembly} (a) Cross-section of the entire power extractor/transfer structure assembly with details of the 3-layer metallic coating of the dielectric tube. (b) Zoom-in view of the contact between the dielectric tube and the rf gasket at the beam exit. (c) Zoom-in view of the contact between the dielectric tube and the tube holder at the beam entrance.}
\end{figure}

The mechanical layout of the X-band dielectric-loaded power extractor is shown in Fig.~\ref{Fig_whole_assembly}. 

The dielectric tube with a metallic coating is fabricated as follows. The dielectric tube, including the uniform and tapered section, was sintered as a single piece with a surface roughness of $\sim$100~nm by grinding. The metallic boundary consists of 3 layers. A chromium buffer layer of $\sim$10~nm was first sputtered onto the dielectric's outer surface followed by a $\sim$100~nm thin copper layer sputtered onto the chromium and finally a thick copper layer of $\sim$75~$\mu$m was electroplated onto the thin copper layer, as illustrated in Fig.~\ref{Fig_whole_assembly}(a). In the theoretical calculation and the CST simulation, the chromium layer is not taken into consideration.

The entire power extractor/transfer structure is assembled as follows. 1) The output coupler and a center flange are connected. The electrical contact and the vacuum seal are ensured by a high power rf gasket in between. 2) This assembly and a copper jacket~\cite{JingIPAC2013} are connected. The electrical contact and vacuum seal are ensured by a regular gasket in between. 3) Three rf absorbers and their supporting rings are loaded into the chamber. The rings have several openings (not shown) to avoid trapped air when pumping. 4) The rf absorbers and the supporting rings are pushed against the center flange by an absorber holder that is screwed onto the chamber. 5) The coated dielectric tube is inserted into the chamber until it comes to a stop against the high power rf gasket. Details of the contact between the tube and the rf gasket are illustrated in Fig.~\ref{Fig_whole_assembly}(b). 6) The tube is held firmly against the rf gasket by a tube holder that screwed onto the absorber holder. The electrical contact between the tube and the tube holder is ensured by tight tolerance and a small step machined on the tube holder, as details illustrated in Fig.~\ref{Fig_whole_assembly}(c).

It should be noted that, during this experiment, the rf absorbers are only used to mechanically support the dielectric tube but they will be used to damp high order modes in future studies. In addition, no cooling is needed since the average output power is on the order of a watt due to the low machine repetition rate (2~Hz) used during the experiment.

\subsection{Cold-test}
\begin{figure}[h!tbp]
	\includegraphics[width=7.5cm]{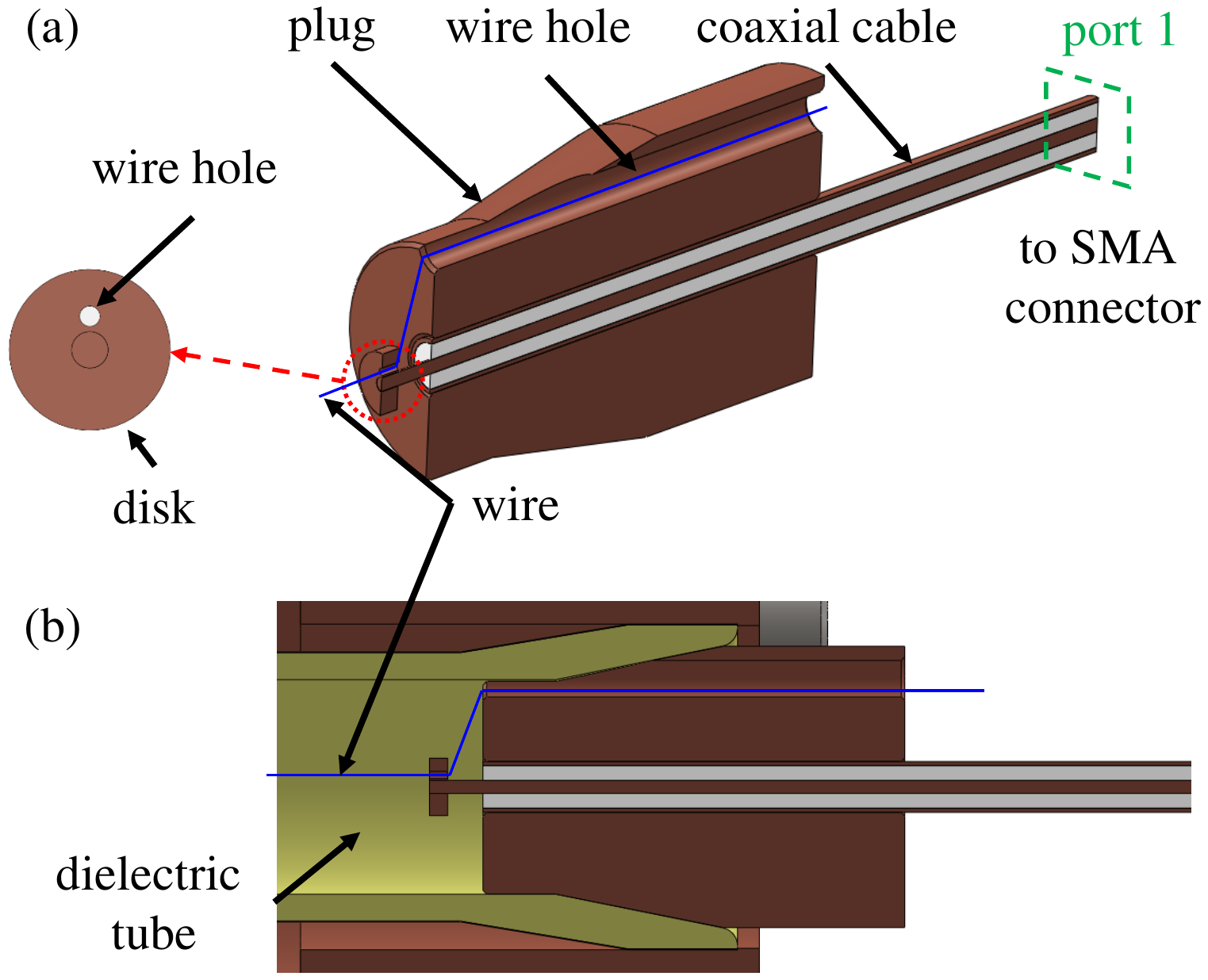}
	\caption{\label{Fig_mode_launcher} (a) The mode launcher used during the cold-test together with a zoomed in view of the disk. (b) The insertion of the mode launcher into the dielectric-loaded power extractor assembly.}
\end{figure}

A mode launcher, as illustrated in Fig.~\ref{Fig_mode_launcher}, was developed for the cold-test since while the structure has an rf coupler at one end, it is open at the other. The launcher converts the TEM mode in the rf cable into the cylindrical TM$_{01}$ mode of the dielectric PETS. 

The assembly procedure of the mode launcher is as follows. 1) A 50~$\Omega$ rigid coaxial rf cable is inserted into a plug that fits tightly against the tapered dielectric section. The electrical contact between the cable and the plug is ensured by tight tolerance. 2) A disk with a small offset hole is welded onto the inner conductor of the rf cable. 3) A nylon wire used for the bead-pull (blue line in Fig.~\ref{Fig_mode_launcher}) is first passed through the hole in the disk and then through one in the plug. 4) The metallic bead is made by tightly wrapping a short piece of metallic tape ($\sim$1~mm long, not shown)  onto the nylon wire.  Note that bead is slightly off-axis  inside the dielectric tube due to the offset hole in the disk. However, the influence of this small offset (0.8~mm) during the bead-pull measurement is negligible according to the CST Microwave Studio simulation.

\subsubsection{S-parameter}
A N5230C network analyzer was used for the cold-test measurements with port 1 connected to the SMA connector of the mode launcher (Port 1 in Fig.~\ref{Fig_mode_launcher}) and port 2 connected to the waveguide output port of the rf coupler (Port 2b in Fig.~\ref{Fig_dielectric_tube}) through a commercial SMA-to-WR90 adapter. The complete setup, including the fully assembled dielectric PETS and the mode launcher, was simulated with CST Microwave Studio. The simulated values of S$_{21}$ and S$_{11}$ at 11.7~GHz are -1.0~dB and -21.3~dB respectively, showing good power coupling and transmission.

\begin{figure}[h!tbp]
	\includegraphics[width=8cm]{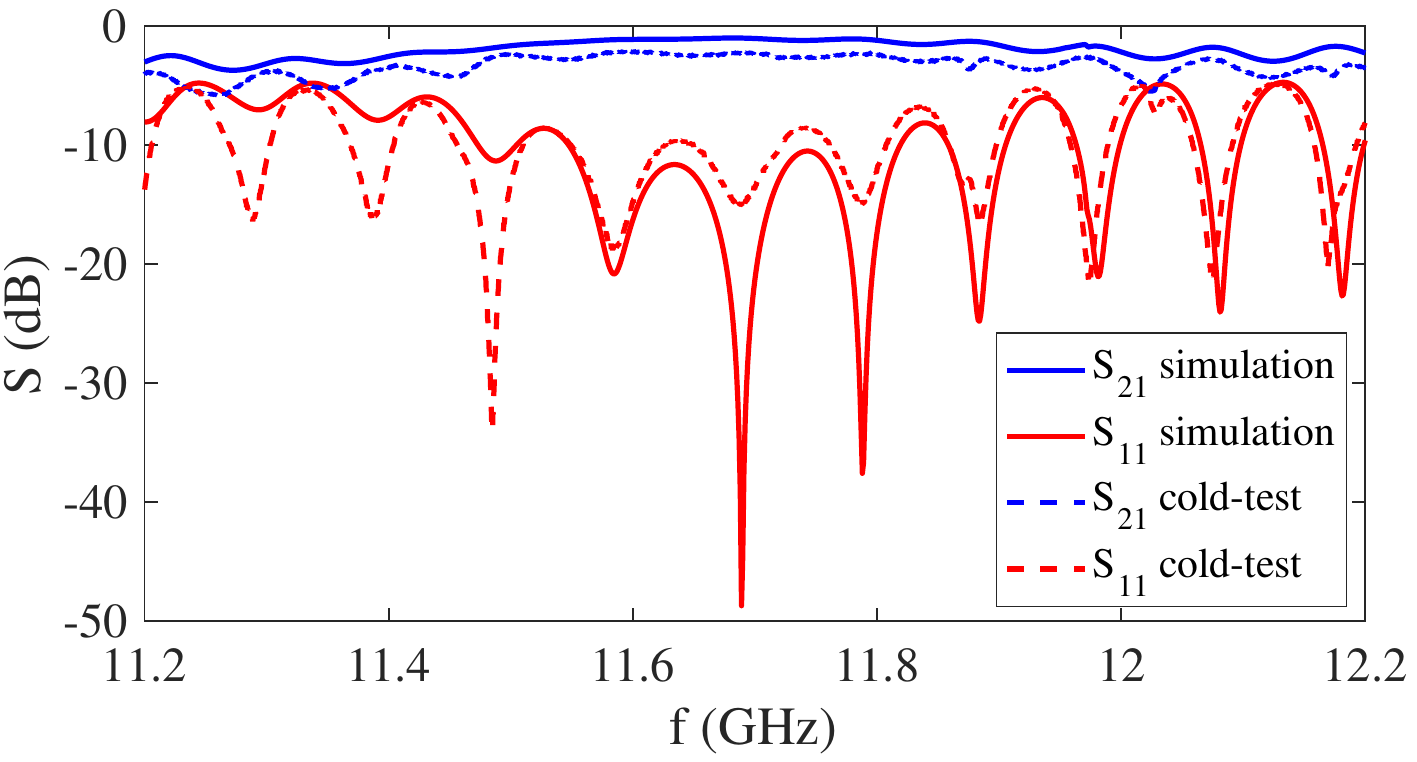}
	\caption{\label{Fig_S_comparison} Comparison of S-parameters from the cold-test measurements and CST simulations of the dielectric-loaded power extractor with the mode launcher.}
\end{figure}

The cold-test measurements are in reasonable agreement with the simulation as indicated by the dashed lines in Fig.~\ref{Fig_S_comparison}. The cold-test values of S$_{21}$ and S$_{11}$ at
11.7~GHz are -2.3~dB and -13.6~dB respectively which are slightly worse than the expected values from the simulation. The low cold-test transmission (2.3-1.0=1.3~dB) can be caused by several factors including: higher rf loss of the structure due to the thin buffer layer of chromium with low conductivity (7.9$\times$10$^{6}$~S/m compared to 5.8$\times$10$^{7}$~S/m of copper), rf losses in the SMA connector of the mode launcher and the SMA-to-WR90 adapter which could not be calibrated, and imperfect electrical contact between the mode launcher and the tube due to machining tolerance.

\subsubsection{Field distribution}
\begin{figure}[h!tbp]
	\includegraphics[width=8cm]{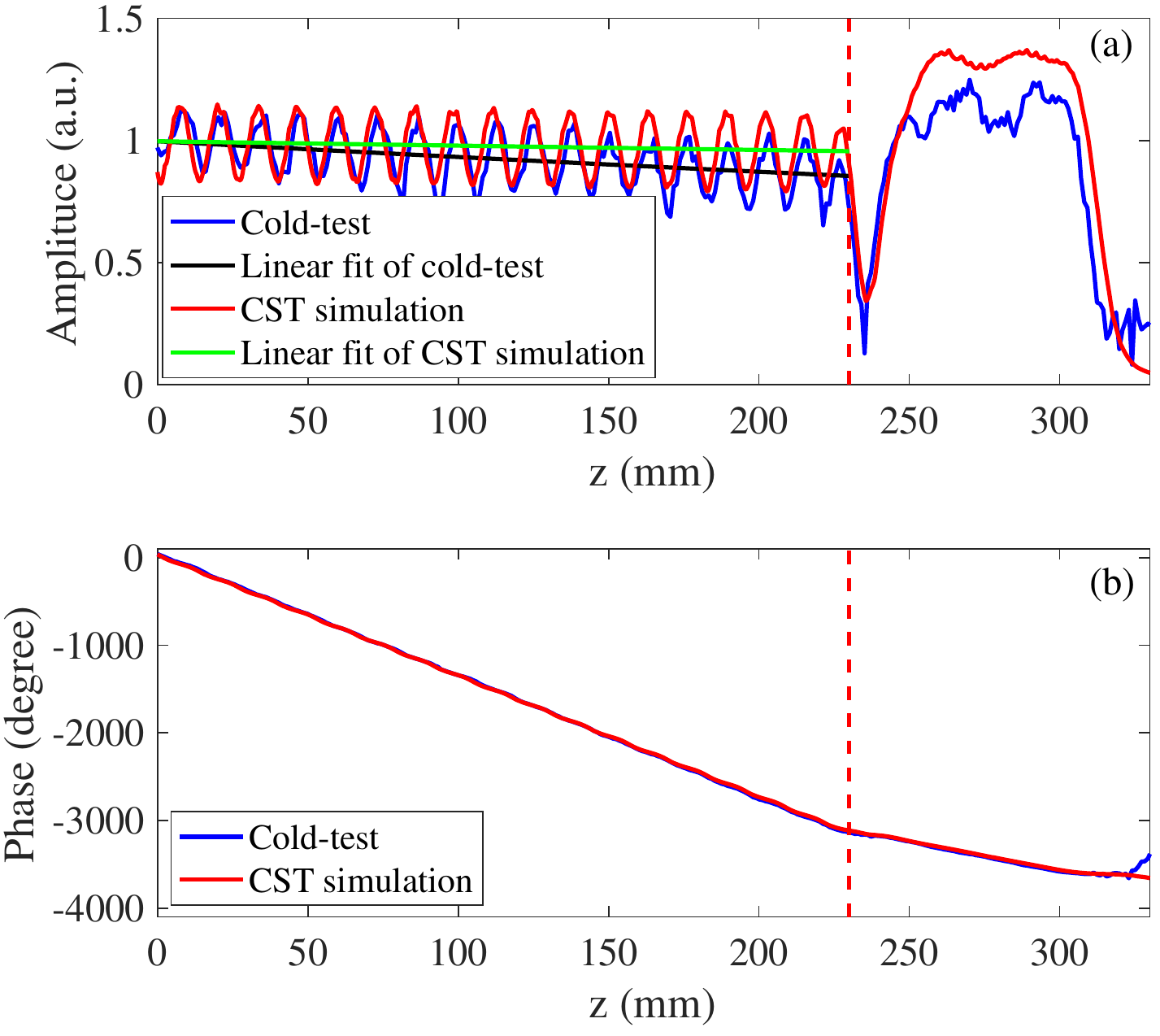}
	\caption{\label{Fig_cold_test_field} The normalized amplitude (a) and phase (b) of the on-axis electric field in the dielectric-loaded power extractor at 11.7~GHz. Due to the insertion of the mode launcher into the dielectric tube, the field at the beam entrance could not be measured. The origin of the horizontal axis has been shifted accordingly. The vertical dashed lines indicate the boundary between the uniform section (left) and the tapered section/coupler (right).}
\end{figure}

The on-axis electric field was measured by the bead-pull method~\cite{JiaruLINAC2010} and is given by the solid blue line in Fig.~\ref{Fig_cold_test_field}. It is in reasonable agreement with the CST Microwave Studio simulation results (solid red lines in Fig.~\ref{Fig_cold_test_field}).

The structure attenuation was calculated from the field amplitude distribution for both the bead-pull measurement and the CST simulation. The field amplitude in the uniform section follows 
\begin{equation}\label{Eqn_attenuation}
E(z)=E_{0}\exp(-\alpha z)\approx E_{0}(1-\alpha z)
\end{equation}
from which the total rf loss of the uniform section is given by $20\log _{10}[\exp(-\alpha L_{s})]$. The oscillation of the field amplitude in the uniform section is caused by the imperfect impedance matching and appears in both the bead-pull measurement and the CST simulation. To calculate $\alpha$, the field oscillation is ignored and the field is approximated by a linear fit.

In Fig.~\ref{Fig_cold_test_field}(a), the linear fits to the CST Microwave Studio simulation results and the cold-test results are presented by the green line and the black line, respectively. For the green line, the fitted $\alpha$ of 0.18~$m^{-1}$ agrees well with the nominal value calculated using the parameters listed in Table.~\ref{Tab_uniform}. The corresponding structure loss is -0.4~dB. For the black line, the fitted $\alpha$ of 0.67~$m^{-1}$ indicates the quality factor of the tube drops from the nominal number of 3393 to $\sim$950. The corresponding structure loss is -1.5~dB. The difference in the rf loss (1.5-0.4=1.1~dB) is consistent with the above S-parameter measurement. The higher structure loss results in a $\sim$12\% reduction of the generated rf power compared to the theoretical calculation and the CST Microwave Studio simulation. According to Eqn.~\ref{Eqn_field_single} and Eqn.~\ref{Eqn_field_multi}, there is no major change to the rf pulse shape caused by the higher loss.

The phase velocity was calculated from the phase distribution for both the bead-pull measurement and the CST simulation. The measured phase of the on-axis electric field follows $\phi (z)=\phi_{0}-\beta z$. Therefore, the propagation constant and the phase velocity follow
\begin{equation}\label{Eqn_phase_measurement}
\left\{
\begin{aligned}
\beta=&\frac{d\phi (z)}{dz}\\
v_{p}=&\frac{2\pi f}{\beta}\\
\end{aligned}
\right.
\end{equation}
where $f$ is the driving frequency in the bead-pull measurement.

\subsubsection{Synchronization frequency}
\begin{figure}[h!tbp]
	\includegraphics[width=8cm]{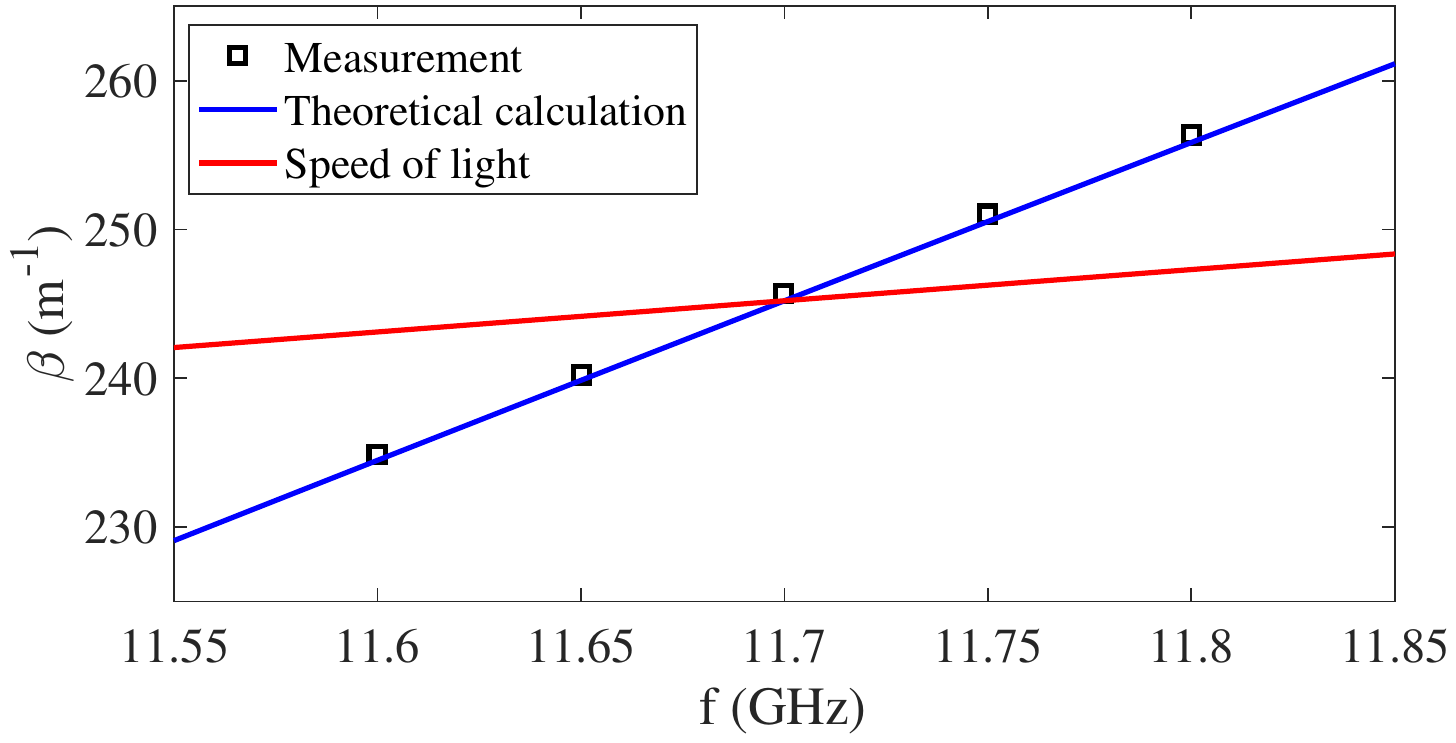}
	\caption{\label{Fig_cold_test_dispersion} Dispersion curve of the dielectric-loaded power extractor.}
\end{figure}

The dispersion curve of the structure was obtained by measuring the phase distribution at various frequencies, as illustrated in Fig.~\ref{Fig_cold_test_dispersion}. The intersection of the dispersion curve (blue) and the line representing the speed of the light (red) indicates $v_{p}=c$ when the beam is continuously decelerated via Cherenkov radiation and generates wakefield at the corresponding frequency. The measurement results agree well with the theoretical calculation of the uniform section by Eqn.~\ref{Eqn_propagation} and Eqn.~\ref{Eqn_diespersion}, showing the wakefield frequency is $\sim$11.69~GHz close to the design value of 11.7~GHz.

To coherently enhance the wakefield from a multi-bunch train generated by the AWA 1.3~GHz rf linac, the interval between each bunch should be $1/11.69 \times 9$=769.9~ps which is 0.7~ps longer than the nominal value. The small offset can be compensated for by launching the bunches in the photocathode rf gun every 360.3$^{\circ}$ instead of every 360$^{\circ}$. The accumulated launching phase difference for the 8-bunch train is $\sim$2$^{\circ}$ which doesn't influence the beam transportation.

\section{\label{cha4}High power test}

\subsection{Experimental setup}
\begin{figure*}[hbtp]
	\centering
	\includegraphics[width=16.5cm]{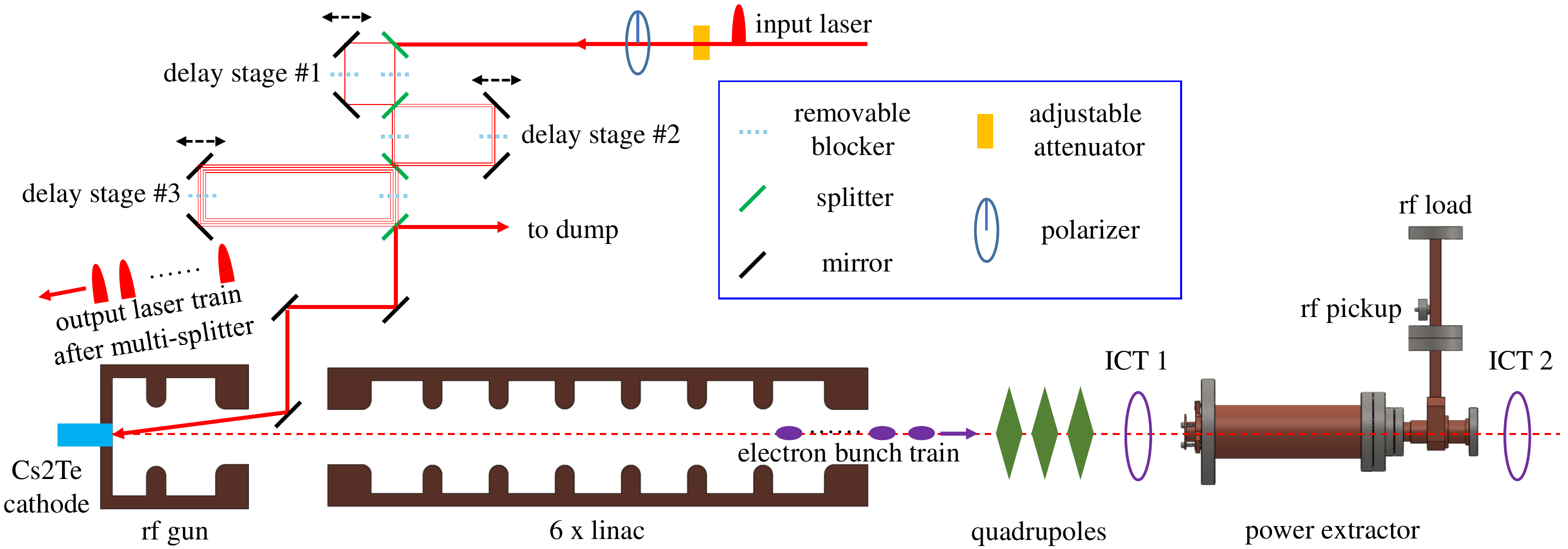}
	\caption{\label{Fig_layout} Schematic layout of the high power experiment at the AWA facility.}
\end{figure*}

The experimental setup used at the AWA facility is illustrated in Fig.~\ref{Fig_layout}. The input laser for the photocathode gun is produced by a Ti:Sapphire laser system and a third harmonic converter to produce a linearly polarized UV (248~nm) pulse with an variable FWHM pulse length of 1.5-10~ps and an energy up to 10~mJ amplified by a KrF excimer~\cite{ManoelIPAC2017}. A UV multi-splitter consisting of four beamsplitters and three delay lines~\cite{NeveuIPAC2016} is used to generate an 8-bunch laser pulse train with $\lambda$ interval. The length of the three delay lines is $\lambda$, 2$\lambda$, and 4$\lambda$ respectively. The value of $\lambda$ is first coarsely set to the 1.3~GHz wavelength (230.6~mm) so that the bunches are launched in every rf bucket of the L-band gun. To fine tune the beam launching phase, the laser pulse interval can be remotely adjusted by moving the position of the mirrors mounted on delay lines. Individual pulses, and  certain pulse combinations can be achieved by the removable laser blockers. The transmission/reflection of the splitters is nominally 50/50 ($\pm$2) percent but also depends on the polarization angle which can be adjusted by a laser polarizer before the multi-splitter. After optimization, the charge balance of the 8-bunch train reached 1.00, 0.95, 1.00, 0.94, 0.91, 0.92, 0.92, and 0.85 in this study. The total charge can be adjusted by a laser attenuator mounted at the entrance of the multi-splitter stage.

The electron beam was produced by the AWA's L-band 1.6-cell rf photocathode gun with a cathode gradient of 62~MV/m and a launching phase of 50$^{\circ}$~\cite{WeiNIMA1998,LianminPRAB2018}. The cesium telluride cathode has a high quantum efficiency of $\sim$10\% which could achieve a maximum charge of $\sim$600~nC~\cite{EricIPAC2015} per pulse train. The transverse laser spot size was set to $\sim$22~mm in order to mitigate the space charge effect so that the electron bunch length can stay short. After the rf gun and the six 7-cell $\pi$-mode L-band standing wave accelerating cavities~\cite{JohnIPAC2010}, the beam was accelerated to 65~MeV, focused by the quadrupoles, and passed through the dielectric-loaded power extractor.

The main diagnostics used during the experiment included two integrating current transformers (ICT)~\cite{ICT} located before and after the structure to measure the charge as well as an rf pickup antenna~\cite{MaomaoIPAC2019} installed at the output waveguide to measure the generated rf power. The signals were recorded by a 4-channel oscilloscope with a 50~GS/s sampling rate and a 20~GHz bandwidth. 

\subsection{Experimental results}
\subsubsection{Fine tuning of the delay stages}
\begin{figure}[h!tbp]
	\includegraphics[width=7.5cm]{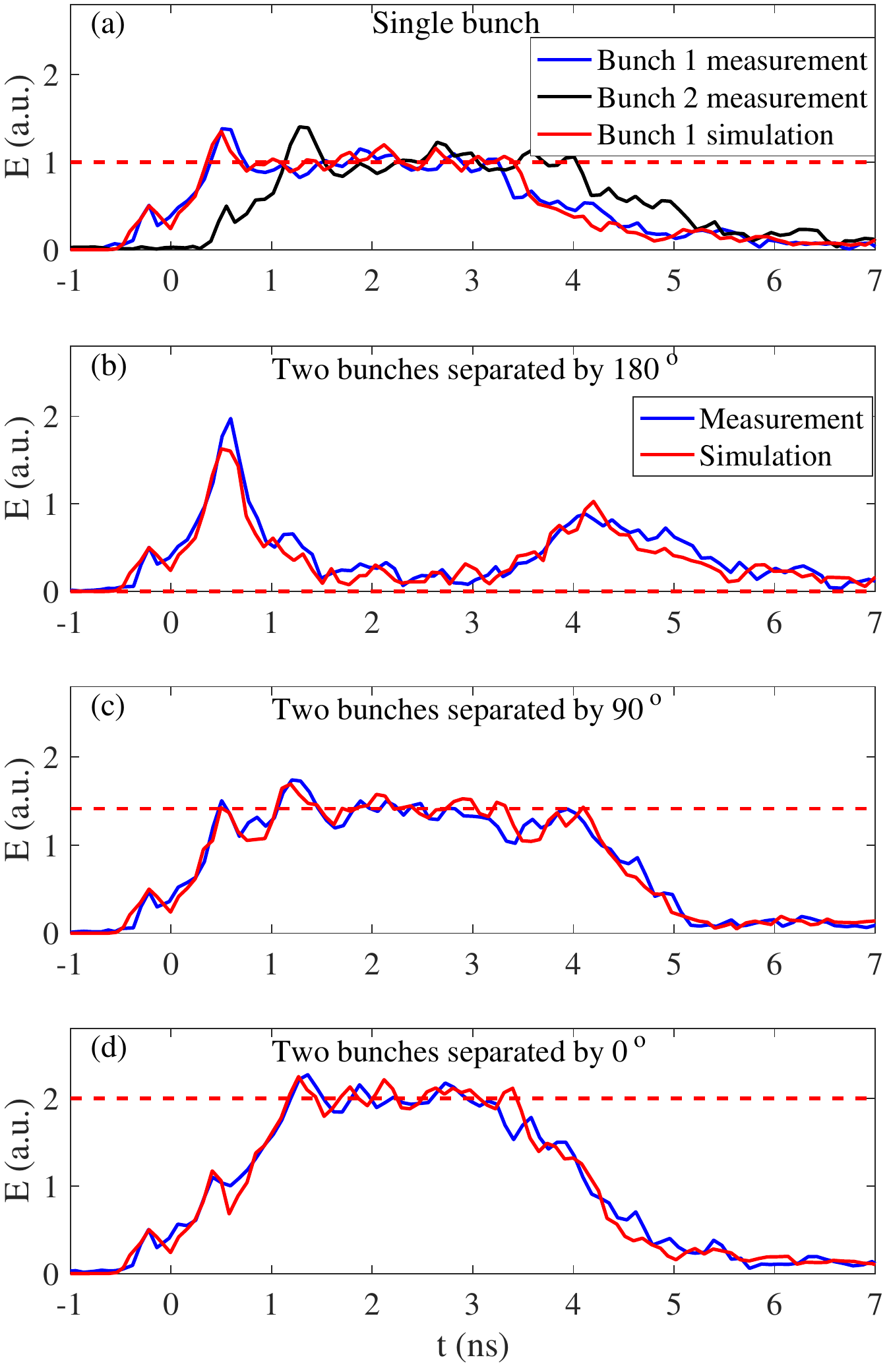}
	\caption{\label{Fig_phase_tuning} Comparison between the rf pulses measured by the pickup and the output port signal by the CST Particle studio simulation when tuning the first stage. The same amplitude normalization factor has been applied to all signals. The red dashed lines in (a-d) denote 1, $2\times \cos(180^{\circ}/2)$, $2\times \cos(90^{\circ}/2)$, and $2\times \cos(0^{\circ}/2)$ respectively. The time, t=0, is set to be the same as that in Fig~\ref{Fig_port_signal}.}
\end{figure}

The launching phases of the bunches were fine tuned using a low charge multi-bunch train by adjusting the delay stages to maximize the generated rf power. Figure~\ref{Fig_phase_tuning} shows an example of tuning the first delay stage using the first two bunches. By ignoring the delay of one L-band period and assuming the same charge, the rf field excited by the two bunches can be expressed as $E_{1}(t)=E\cos(\omega t)$ and $E_{2}(t)=E\cos(\omega t +\bigtriangleup \theta)$, where $E$ is the field amplitude and $\bigtriangleup \theta$ is the relative phase difference. The super-posited field amplitude is therefore equal to $E_{12}=E \times 2\cos(\bigtriangleup \theta/2)$. When moving the delay stage to change $\bigtriangleup \theta$, the experimental results agree very well with the CST Particle studio simulation in which $E_{2}$ is duplicated from $E_{1}$ with a phase shift. Using the same method, the second and the third delay stages can be tuned with the first two 2-bunch train (bunch No.1-2 and bunch No.3-4) and the two 4-bunch train (bunch No.1-4 and bunch No.5-8), respectively. The accuracy of the tuning is better than 1$^{\circ}$ in L-band or 9$^{\circ}$ in X-band. 

\subsubsection{High power generation}
\begin{figure}[h!tbp]
	\includegraphics[width=7.5cm]{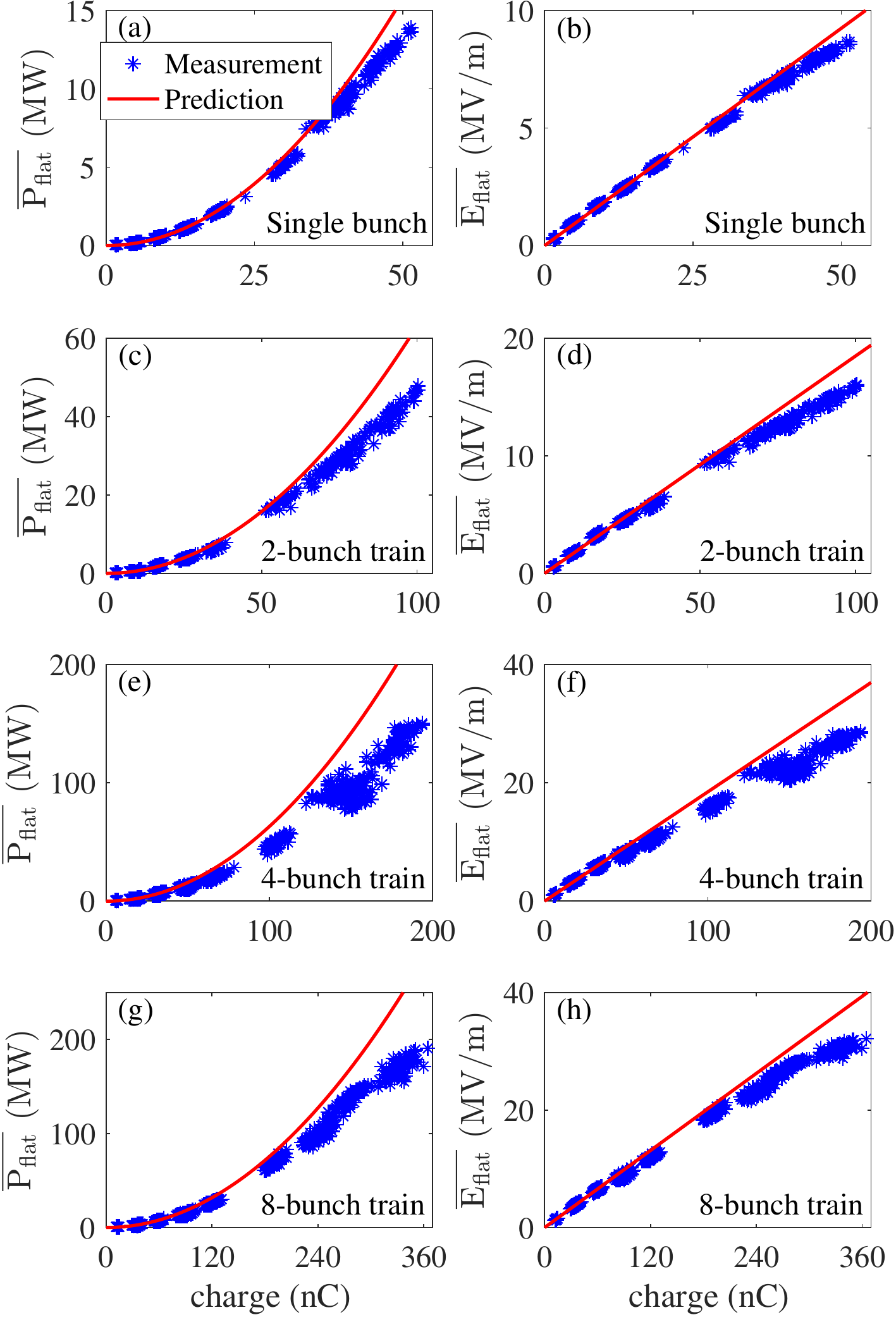}
	\caption{\label{Fig_power_gradient} Comparison of the rf power (left) and the wakefield gradient (right) of the flat-top between the experimental result and the theoretical prediction. In the prediction, the form factor is fixed at 0.96; the quality factor of the uniform section is set to 950 according to the cold-test results; and the realistic charge balance is used.}
\end{figure}

The duration and the power level of the generated rf pulses were gradually increased by increasing the number and the charge of bunches in the drive train after setting the delay stages. This process is similar to rf conditioning in externally rf-driven metallic high-gradient structures where the power is increased to raise the gradient~\cite{WalterPRAB2017,XiaoweiPRAB2017}. The single bunch (bunch No.1), the 2-bunch train (bunch No.1-2), the 4-bunch train (bunch No.1-4), and the 8-bunch train (bunch No.1-8) were applied successively to increase the flat-top duration. With each bunch train configuration, the charge was slowly increased from zero to the maximum value (50~nC for the single bunch, 100~nC for the 2-bunch train, 195~nC for the 4-bunch train, and 360~nC for the 8-bunch train) to increase the power level, as illustrated in Fig.~\ref{Fig_power_gradient}. The solenoids along the beamline (not shown in Fig. ~\ref{Fig_layout}) and the quadrupoles were adjusted to ensure $\sim$100\% charge transmission through the structure.

\begin{figure}[h!tbp]
	\includegraphics[width=7.5cm]{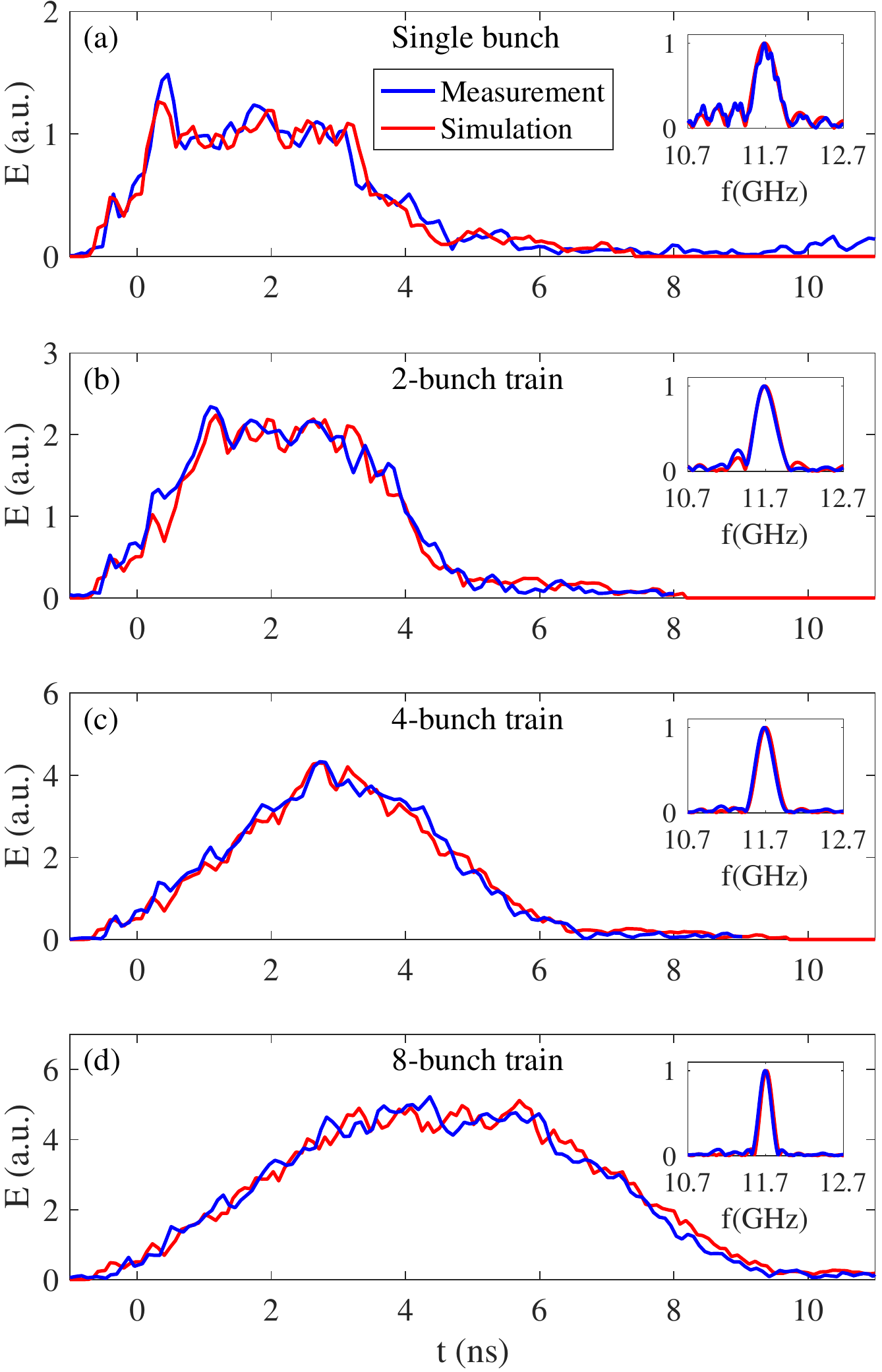}
	\caption{\label{Fig_port_signal_measurement} Comparison of the field envelop and the spectrum (inset) of the coupler output rf signal between the experimental measurement and the theoretical prediction. The time, t=0, is set to be the same as that in Fig~\ref{Fig_port_signal}.}
\end{figure}

When driven by the 8-bunch train, with a total charge of 360~nC, the structure successfully generated $\sim$200~MW, $\sim$3~ns flat-top rf pulses. The corresponding wakefield gradient derived by Eqn.~\ref{Eqn_power_flat_top} was $\sim$32~MV/m. The maximum $U_{loss}$ of the bunches was less than 6~MeV according to Eqn.~\ref{Eqn_energyloss_multi}.

rf breakdown can lead to an abrupt power drop within the transmitted/generated rf pulses~\cite{WalterPRAB2017,XiaoweiPRAB2017} but this was not observed during the experiment. Figure.~\ref{Fig_port_signal_measurement} illustrates the comparison between the simulated rf pulse shapes and the measured ones with the highest charge. The good agreement indicates that there was no structure damage during the measurement.

\subsection{Data analysis}
Beam dynamics simulations with the General Particle Tracer (GPT) code~\cite{GPTmanual} were conducted to understand the discrepancy between the measured rf power and the predicted value when pushing the power. The rms bunch length was simulated from the cathode to the power extractor entrance with similar beam conditions and beamline settings as those in the experiment. The space charge as well as the reduced field in the rf gun/linacs due to the beam loading effect~\cite{JohnIPAC2010} were taken into consideration.

\subsubsection{Single bunch}
The derived $\overline{E_{flat}}$ from the single bunch measurement is linearly proportional to the charge in the regime where the charge is less than 20~nC, as illustrated in Fig.~\ref{Fig_power_gradient}(b). Therefore, the form factor can be fitted to be 0.96 from Eqn.~\ref{Eqn_field_flat_top} and Eqn.~\ref{Eqn_field_single}. The corresponding rms bunch length is 1.2~mm, which agrees well with the GPT simulated value of 1.1~mm. 

The discrepancy of $\overline{E_{flat}}$ between the measurement and the prediction with increasing charge is suspected to be caused by a decrease of the form factor.  This is a result from a longer bunch length due to space charge forces and beam loading effects.  The form factor obtained from a fit of the high charge data is $\sim$7~\% lower which corresponds to an rms bunch length of 2.0~mm; in reasonable agreement with the GPT simulated value of 1.6~mm. The difference might be caused by the non-Gaussian distribution of the drive beam longitudinal profile in experiment.

\subsubsection{Bunch train}
The aforementioned form factor of 0.96 is used to calculate the generated rf power when driven by multi-bunch trains. The measurement results show good agreement with the predictions with low charges ($\leq$20~nC per bunch), as illustrated in Fig.~\ref{Fig_power_gradient}(c-h). 

The deviation between the measurement results and the theoretical predictions observed at high charges is similar to the single bunch case. With the highest charge of the 8-bunch train, the measured rf power is $\sim$70~\% of the prediction. It's worth mentioning that the percentage is close to that reported with an X-band metallic power extractor tested at AWA under similar beam conditions~\cite{JingNIMA2018}. Therefore, the low power generation is more likely caused by factors other than the structure itself. The rms bunch length of the last bunch is simulated to be 1.9~mm, slightly longer than the first one due to the beam loading effect in the rf gun and linacs. The average bunch length of the 8-bunch train is $\sim$1.7~mm, which leads to a form factor of 0.91 and a power reduction of 10~\%. If we also consider non-ideal delay stages tuning (9$^{\circ}$ in X-band between adjacent bunches), the generated power could be reduced by another 12~\% due to imperfect wakefield coherent superposition. These two factors could lead to a maximum of $\sim$21~\% lower generated power at high charge. The remaining difference between the analysis and the experimental result might be caused by non-Gaussian longitudinal distribution, multipacting of the rf pickup~\cite{MaomaoIPAC2019}, and other unknown factors. A detailed study of the time structure of the drive bunches was not performed in this experiment but will be investigated in the future. 

\section{\label{cha5}Future study}
In this section we present the design study of a future dielectric-loaded power extractor in order to maximize the generated rf power beyond the 200~MW level achieved in this study. This is done by optimizing the design of the dielectric-loaded power extractor while assuming similar beam parameters to the ones that have been achieved in this study. The drive beam is assumed to contain an 8-bunch train with $\sigma _{z}$ of 1.2~mm, $T_{b}$ of 769.2~ps, $Q_{b}$ of 50~nC, and prefect charge balance. This represents a total drive train charge of 400~nC which is below the 600~nC level previously demonstrated at AWA~\cite{EricIPAC2015}. $L_{st}$ and $v_{g}$ are fixed at 26~cm and 0.184~c respectively so that $N_{rise}=4$ and $t_{flat}$=3.1~ns. The frequency choice of the future structure is considered for L-band harmonics starting from C-band (7.8~GHz) to Ka-band (32.5~GHz).  In addition, several inner radii are used for each structure design. For each combination of $\omega _{0}$, $v_{g}$, and $a$, $\epsilon _{r}$ is calculated from Eqn.~\ref{Eqn_propagation} and Eqn.~\ref{Eqn_diespersion}, with the upper limit set to 20 for the ease to obtain low rf loss dielectric materials.

\begin{figure}[h!tbp]
	\includegraphics[width=8cm]{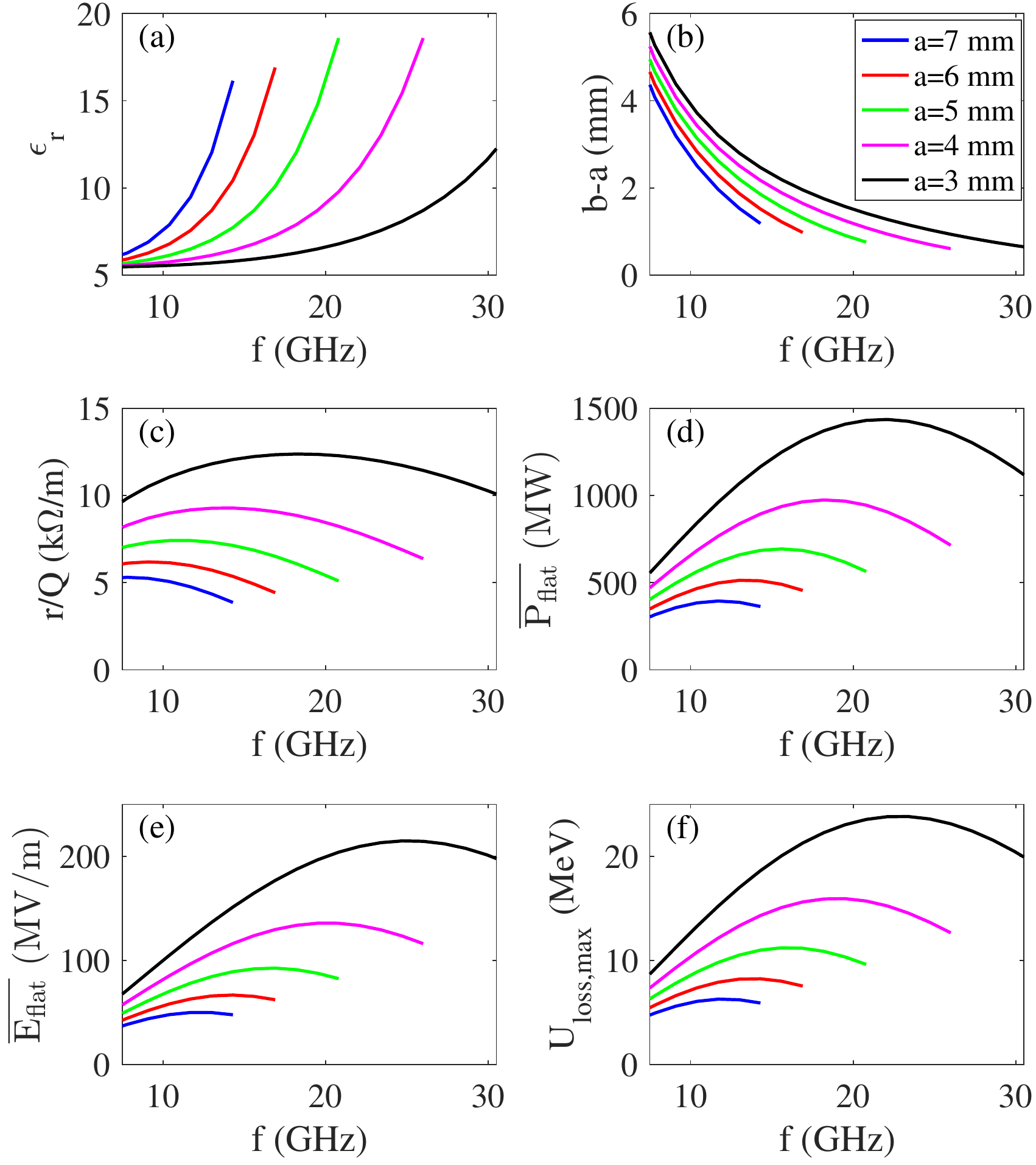}
	\caption{\label{Fig_towards_highpower} Electromagnetic and wakefield properties of the uniform section as a function of the operation frequency with various inner radii. (a) Dielectric constant. (b) Dielectric thickness. (c) $r/Q$. (d) $\overline{P_{flat}}$. (e) $\overline{E_{flat}}$. (f) Maximum $U_{loss}$.}
\end{figure}

The optimization results of the uniform section are shown in Fig.~\ref{Fig_towards_highpower}. The maximum achievable rf power increases when reducing the inner radius and the corresponding operation frequency shifts to higher values. When $a\leq$4~mm, gigawatt level can be obtained around 20~GHz. In this regime, the dielectric-loaded power extractors are feasible in fabrication: the required low rf loss dielectric materials are commercially available, such as Forsterite ($\epsilon _{r}$=6.64)~\cite{JingRAST2016} and alumina; and the wall thickness is $\sim$1.5~mm. The structures are also physically achievable: the corresponding gradient is well below the GV/m level reported in previous rf breakdown tests~\cite{ThompsonPRL2008} and CWA experiments~\cite{OSheaNatureComm2016}; and the maximum $U_{loss}$ of the bunches is $\sim$1/3 of the AWA beam energy. The detailed parameters of an possible design are provided in Table~\ref{Tab_uniform_highpower}.

\begin{table}[h!tbp]
	\caption{\label{Tab_uniform_highpower}Parameters of the uniform section of a future gigawatt dielectric-loaded power extractor using the AWA drive beam conditions.}
	\begin{ruledtabular}
		\begin{tabular}{l c}
			Parameter&Value\\
			\hline
			Material&Forsterite\\
			Dielectric constant&6.64\\
			Loss tangent&$1 \times 10^{-4}$\\
			Frequency&19.5~GHz (15 $\times$ 1.3~GHz)\\
			Inner radius&3.1~mm\\
			Outer radius&4.6~mm\\
			Length&26~cm\\
			Quality factor&3177\\
			$r/Q$&11.93~k$\Omega$/m\\
			Group velocity&0.184~c\\
			$E_{max,surface}/E_{axis}$&1\\
			Flat-top duration&3.1~ns\\
			$\overline{P_{flat}}$&1.36~GW\\	
			$\overline{E_{flat}}$&190~MV/m\\	
			Maximum $U_{loss}$&22.2~MeV\\			
		\end{tabular}
	\end{ruledtabular}
\end{table}

The major challenge in using this future structure will be the transmission of the high charge drive beam through the structure. Increasing transmission requires several efforts: improve the beam quality via low emittance and short bunch length~\cite{Futuresources2016}, implement beam breakup control~\cite{LiPRST2014}, add HOMs damping~\cite{JingIPAC2013}, etc.

The other challenges include design of rf couplers and loads for gigawatt power level, improvement in the coating technology to reduce structure loss, rf breakdown and multipacting study for HOM damped structures, time structure characterization of the bunch train, etc.

\section{\label{cha6}Summary}
The dielectric-loaded structure is a promising candidate, as both the power extractor and the accelerator, for the short-pulse two-beam acceleration scheme having the advantage of simple geometry, low fabrication cost, high group velocity with reasonable shunt impedance, and the potential to withstand GV/m gradient. In order to demonstrate rf power generation beyond 100~MW, an X-band 11.7~GHz dielectric-loaded power extractor has been developed and tested at the Argonne Wakefield Accelerator facility. The detailed development process including design, simulation, fabrication, cold-test, and high power test, are presented. Driven by 8-bunch trains with a total charge of 360~nC, the structure successfully generated $\sim$200~MW, $\sim$3~ns flat-top rf pulses without rf breakdown. Plans for a future structure optimization shows that gigawatt power level is feasible at the AWA facility by using higher frequency structures with smaller apertures. The work benefits the research of power extractors in the SWFA scheme.

\begin{acknowledgments}
The work at AWA is funded through the U.S. Department of Energy Office of Science under Contract No. DE-AC02-06CH11357. The work by Euclid Techlabs LLC is funded through the U.S. Department of Energy under SBIR Contract No. DE-SC00011299. We would like to thank Cho Ng, Zenghai Li, Liling Xiao, and Lixin Ge at SLAC for the ACE3P simulation. We would also like to thank Kiran Kumar Kovi and Gongxiaohui Chen from Euclid Techlabs LLC as well as Christopher Marshall and Michael Schwartz from the Chemical Sciences and Engineering Division at Argonne National Laboratory for the metallic coating of the dielectric tube.
\end{acknowledgments}

\bibliography{XDPETS_ref}

\end{document}